\documentclass[%
 showpacs,
 amsmath,amssymb,
]{revtex4-1}

\usepackage{graphicx}
\usepackage{dcolumn}
\usepackage{bm}

\usepackage{CJK}

\usepackage{color}
\usepackage{amssymb}
\usepackage{amsfonts}

\usepackage[colorlinks,urlcolor=blue,linkcolor=blue,citecolor=blue,anchorcolor=blue]{hyperref}

\begin{document}

\preprint{APS/123-QED}

\title{Photonic Floquet Topological Insulators in Atomic Ensembles}

\author{Yiqi Zhang$^1$}
\email{zhangyiqi@mail.xjtu.edu.cn}
\author{Zhenkun Wu$^1$}
\author{Milivoj R. Beli\'c$^{2}$}
\author{Huaibin Zheng$^1$}
\author{Zhiguo Wang$^1$}
\author{Min Xiao$^{3,4}$}
\author{Yanpeng Zhang$^{1}$}

\affiliation{%
 $^1$Key Laboratory for Physical Electronics and Devices of the Ministry of Education \& Shaanxi Key Lab of Information Photonic Technique,
Xi'an Jiaotong University, Xi'an 710049, China \\
$^2$Science Program, Texas A\&M University at Qatar, P.O. Box 23874 Doha, Qatar \\
$^3$Department of Physics, University of Arkansas, Fayetteville, Arkansas, 72701, USA \\
$^4$National Laboratory of Solid State Microstructures and School of Physics, Nanjing University, Nanjing 210093, China
}%

\date{\today}

\begin{abstract}
  \noindent
  We {demonstrate} the feasibility of realizing a photonic Floquet topological insulator (PFTIs) in an atomic ensemble.
  The interference of three coupling fields will split energy levels periodically,
  to form a periodic refractive index structure with honeycomb profile
  that can be adjusted by different frequency detunings and intensities of the coupling fields.
  { This in turn will} affect the appearance of Dirac cones in {the} momentum space.
  When the honeycomb lattice sites are helically ordered along the propagation direction, gaps open at Dirac points, and
  one obtains a PFTI in { an atomic vapor. An }obliquely incident beam will be able to move along the zigzag edge { of the lattice}
  without scattering energy into the PFTI, due to the confinement of edge states.
  The appearance of Dirac cones and { the formation of} PFTI
  can be shut {down} by the third-order nonlinear susceptibility and opened up by the fifth-order one.
\end{abstract}

\maketitle

%
\section{Introduction}

Recently, topological insulators (TIs) have attracted much attention, since TI,
as a new phase of matter, only allows conducting electrons to exist on the surfaces,
and the moving electrons {are not} affected by defects or disorder \cite{kane_prl_2005, hsieh_nature_2008, hasan_rmp_2010}.
There are edge states {in} TI that lie in a bulk energy gap in {the} momentum space and are spatially localized on the boundaries of TI.
The edge states are predicted to be useful in performing quantum computations \cite{qi_rmp_2011}.
TIs, as well as some graphene-based structures, have also found potential applications in optical modulators \cite{liu_nature_2011, yu_lpor_2013} and optical diodes \cite{liang_prl_2013}.
Photonic topological insulators (PTIs), fabricated by using metamaterials \cite{khanikaev_nm_2012} or helical waveguides \cite{rechtsman_nature_2013},
can break the time-reversal symmetry and lead to one-way edge states, which are robust against defects.

So far, { research on PTIs has been mostly} based on graphene-like structures.
A honeycomb lattice \cite{peleg_prl_2007, ablowitz_pra_2009, lee_pra_2009, bahat-treidel_prl_2010, bahat-treidel_pra_2011}
also exhibits certain graphene-like properties
and can be obtained by using the femtosecond laser writing technique or the three-beam interference method \cite{terhalle_prl_2008, plotnik_nm_2014}.
The first method is valid only in solid materials,
{whereas} the second method can be used in both solid and gas materials \cite{zhang_prl_2011, zhangyanpeng_oe_2010, zhang_lpl_2013}.
We note that the three-beam interference will generally induce a hexagonal lattice instead of the honeycomb lattice.
However, the corresponding refractive index {modulation} will exhibit honeycomb profile in a saturable nonlinear medium or an atomic vapor.

The interference pattern ({in the form of} hexagonal lattice) produced by the three-beam interference
will exhibit many pairs of singularities,
and the band structure of the corresponding refractive index change (in the form of honeycomb lattice)
will feature conical singularities at the corners of the first Brillouin zone.
In an atomic ({e.g.} rubidium) vapor, when the three-beam interference pattern serves as the dressing field, the dressed atomic system will exhibit controllable optical properties,
which were extensively investigated in the past decade \cite{artoni_prl_2006,zhang_ieee_2012, wang_prl_2013}.

In this paper, we investigate the creation of photonic Floquet topological insulators (PFTIs) in an atomic ensemble,
which, to the best of our knowledge, has {not been studied before.}
Even though there is a related work done in ultracold fermionic atoms \cite{jotzu_nature_2014},
the topic {of} this work is quite different.
{ It opens the possibility of generating PFTIs in atomic
vapors, under physical mechanisms that are vastly different
from the ones in solid state materials.
One of the main advantages
in utilizing atomic media for the generation of PFTIs}
is that many interesting topological properties can be easily controlled through adjusting frequency detunings
and powers of the coupling fields, as well as higher-order nonlinear susceptibilities,
{which are readily available in} multi-level atomic systems.

{The paper is organized in the following manner. In Sec.
\ref{model} the model is introduced, based on an inverted Y-type
energy level system in an atomic vapor. Section \ref{pbg} deals with
the photonic band structure and Sec. \ref{PFTI} with the generation
of the photonic Floquet topological insulator in the chosen
system. Section \ref{discussion} provides a discussion of results and Sec.
\ref{conclude} brings conclusion.}

\section{The model}{\label{model}}

\begin{figure}[htbp]
\centering
  \includegraphics[width=0.5\columnwidth]{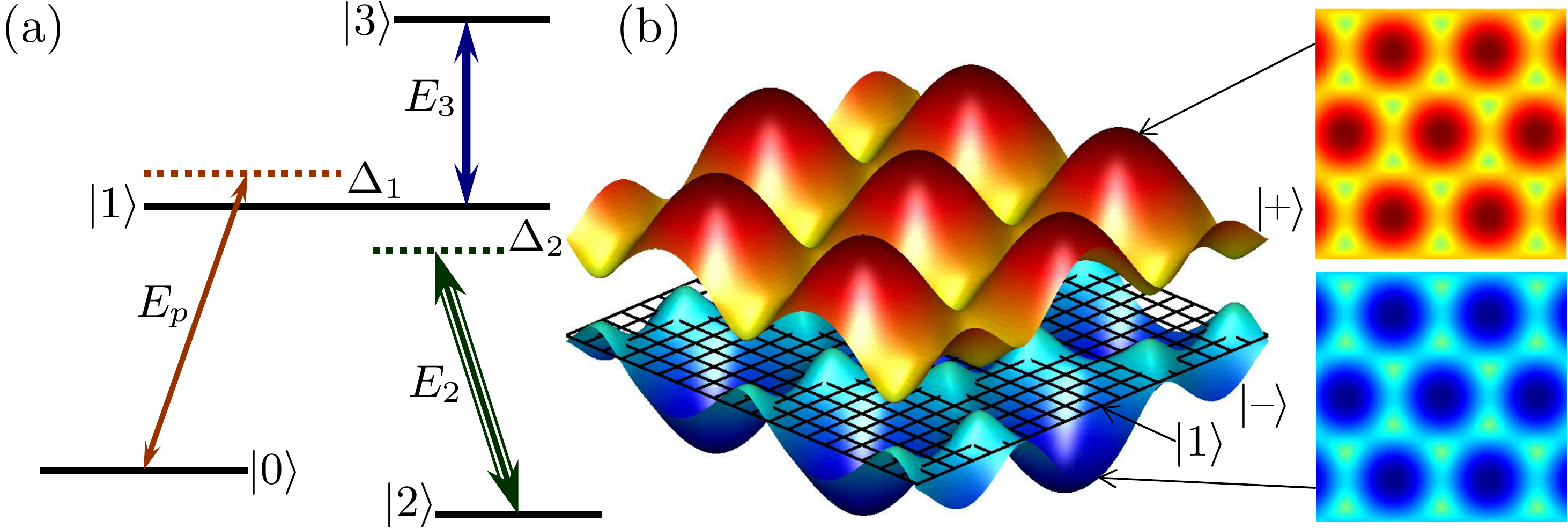}
  \caption{(Color online) (a) Inverted Y-type energy system level scheme with $5S_{1/2}\,(F=3) $ ($|0\rangle $),
  $5P_{3/2} $ ($|1\rangle $), $5S_{1/2}\,(F=2) $ ($|2\rangle $), and $5D_{3/2} $ ($|3\rangle $) of $^{85}$Rb atoms \cite{steck_alkali}.
  (b) Energy level splitting due to the three-beam interference pattern with $\Delta_2=0$.
  The two right panels are the top-view of the two split sublevels.}
  \label{fig1}
\end{figure}

We consider an inverted Y-type electromagnetically induced transparency (EIT) {atomic} system,
as shown in Fig. \ref{fig1}(a), in which $E_p$ probes the transition $|0\rangle \to |1\rangle $,
the coupling field $E_2$ drives the transition $|1\rangle \to |2\rangle $,
and the controlling field $E_3$ connects $|1\rangle \to |3\rangle $.
The proposed scheme is for an atomic system in regular cold magneto-optical trap.
If three coupling fields are used with the same frequency and launched along the same direction $z$,
they will interfere with each other \cite{pang_jpsj_2011,liyongyao_pra_2010,liyongyao_pra_2012},
to form a two-dimensional hexagonal lattice interference pattern in the transverse $xy$ plane.
The resulting Rabi frequency of such an optically induced interference pattern can be written as
\begin{equation}\label{eq1}
G =  \sum_{i = 1}^3 {G _2} \exp [ik_2(x\cos {\theta _i} + y\sin {\theta _i})],
\end{equation}
where $\theta_i=[0,\,2\pi/3,\,4\pi/3] $ are the relative phases of the three laser beams \cite{terhalle_prl_2008},
$k_2 $ is the wavenumber of the coupling fields, and
$G_2 $ represents Rabi frequency of the coupling field, with  $G_2=\wp_{12} E_2/\hbar$,
where $\wp_{12}$ is the electric dipole moment.
Level $|1\rangle $ can be dressed by the coupling fields \cite{wu_josab_2008} and
split into two sublevels $|+\rangle $ and $|-\rangle $ having modified eigenfrequencies
$\ell_{|\pm\rangle} = -\Delta_2/2 \pm \sqrt{\Delta_2{}^2/4+|G|^2} $
with $|G|^2=|G_2|^2 [ 4\cos (3k_2 x/2) \cos(\sqrt{3} k_2 y/2) +  2\cos(\sqrt{3} k_2 y)+3]  $.
Since the three coupling fields interfere with each other to form a periodic interference pattern,
the sublevels $|\pm\rangle $ will be periodic, as shown in Fig. \ref{fig1}(b) and the inset panels,
in which the grid represents the original level $|1\rangle $.

\begin{figure*}[htbp]
\centering
  \includegraphics[width=0.7\textwidth]{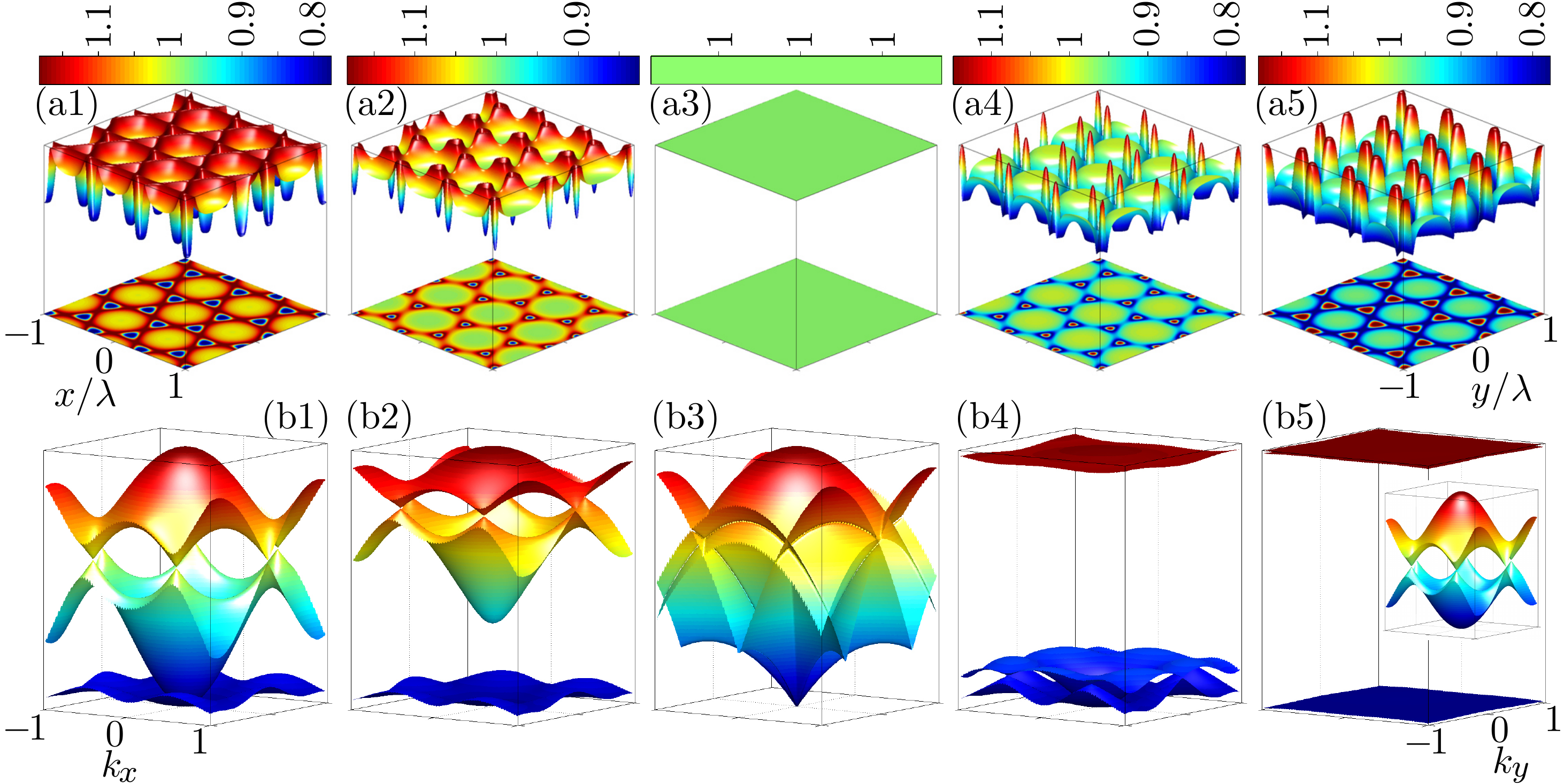}
  \caption{(Color online) (a1)-{(a5) Refractive index change} $\Delta n(x,\,y,\,z) $ with $\Delta_1=-\Delta_2=-10 $ MHz (a1), $\Delta_1=-\Delta_2=-5 $ MHz (a2),
  $\Delta_1=\Delta_2=0 $ (a3), $\Delta_1=-\Delta_2=5 $ MHz (a4), $\Delta_1=-\Delta_2=10 $ MHz (a5) and $V_0=60 $ under the same color scale.
  (b1)-(b5) The corresponding PBG structures, in which the first three bands are shown.
  The inset in (b5) is zoomed-in plot of the first two bands.}
  \label{fig2}
\end{figure*}

Since the energy level is dressed to be periodic,
the susceptibility will also have the same periodicity,
which can be written as \cite{wu_pra_2013, paredes_prl_2014}
\begin{align}\label{eq2}
\chi_p (x,\,y)= {\chi_p^{(1)}} + {\chi_p^{(3)}|E_2|^2} + {\chi_p^{(5)}|E_2|^4},
\end{align}
when {only} the first three (linear and nonlinear) susceptibilities are considered.
In Eq. (\ref{fig2}), \[\chi_p^{(1)}={iN\wp _{10}^2}/\{ \hbar \epsilon_0 [(d_{10} + |G|^2/d_{20})] \},\]
\[\chi_p^{(3)}|E_2|^2 = {-iN\wp _{10}^2 G_2^2}/\{\hbar \epsilon_0 [d_{20} (d_{10} + |G|^2/d_{20})^2]\}, \]
and \[\chi_p^{(5)}|E_2|^4 = {iN\wp _{10}^2 G_2^4}/\{\hbar \epsilon_0 [d_{20}^2 (d_{10} + |G|^2/d_{20})^3]\},\]
with $N$ being the atomic density, $\wp_{10} $ the electric dipole moment,
{and $d_{10}={\Gamma _{10}} + i{\Delta _1} $ and $d_{20}=\Gamma _{20} + i({\Delta _1} - {\Delta _2})$
the complex relaxation rates. Here}
$\Gamma_{ij} $ is the decay rate between $|i\rangle $ and $|j\rangle $,
$\Delta_1=\Omega_{10}-\omega_p $ and  $\Delta_2=\Omega_{12}-\omega_2 $ are the frequency detunings,
$\Omega_{ij} $ is the transition frequency between $|i\rangle $ and $|j\rangle $,
and $\omega_p $ ($\omega_2 $) is the frequency of the probe (coupling) field.
We would like to note that the optical lattice structures in atomic systems are quite stable,
as {long} as the laser beams forming the optical lattice are stable;
small fluctuations should not affect our main results.

The paraxial propagation of the probe beam in the medium can be described by the Schr\"odinger-like equation \cite{efremidis_prl_2003, rechtsman_nature_2013, plotnik_nm_2014}:
\begin{align}\label{eq4}
i\frac{{\partial \psi }}{{\partial z}} =  - \frac{1}{{2{k_0}}}{\nabla ^2}\psi
- V_0 \frac{{{k_0}\Delta n(x,y)}}{{{n_0}}}\psi ,
\end{align}
where $\psi(x,\,y,\,z) $ is the electric field envelope of the probe beam, and
$\nabla^2 = \partial/\partial x^2 +  \partial/\partial y^2 $ {is the transverse Laplacian}.
$\Delta n(x,\,y,\,z) $ in Eq. (\ref{eq4}) is the ``effective potential'' induced by the coupling fields according to Eqs. (\ref{eq1}) and (\ref{eq2}),
$V_0$ is the potential depth, and $k_0=2\pi n_0/\lambda $ is the wavenumber of the probe beam, with $\lambda$ being the wavelength.
In an atomic vapor, the ambient refractive index (RI) is $n_0=1 $, and the wavelength is assumed to be $\lambda=780~\rm nm$.
In Eq. (\ref{eq4}), the effective potential can then be written as
$ \Delta n (x,\,y)= \sqrt {1 + {\rm Re} \{\chi_p(x,\,y)\}} \approx
\delta n_0 + \delta n_1 [2\cos(3k_2 x/2) \cos(\sqrt{3} k_2 y/2) +  \cos(\sqrt{3} k_2 y)] $
if only the first-order susceptibility $\chi_p^{(1)}$ is considered,
where $\delta n_0 \approx \sqrt{{\rm Re} \{1+\eta\} } $ is the spatially uniform RI;
$\delta n_1 \approx -{\rm Re} \{\eta \xi\} / (2\delta n_0) $ is the coefficient for the spatially varying terms for the modulated RI,
with \[\eta=iN \wp_{10}^2 d_{20} /[\hbar \epsilon_0 (d_{10} d_{20} + 3|G_2|^2)] \] and \[\xi \approx 2|G_2|^2 /(d_{10}d_{20}  + 3|G_2|^2). \]

\section{Photonic band gap structure}\label{pbg}

In Figs. \ref{fig2}(a1) and \ref{fig2}(b1), we display schematic diagrams of the RI modulation
and the corresponding photonic band gap (PBG) structure
according to the plane wave expansion method.
We use the plane wave expansion method to obtain band structure of different interference patterns,
because this method takes into account all the {necessary} parameters which affect the RI modulation.
According to previous literature \cite{rechtsman_nature_2013,plotnik_nm_2014,szameit_pra_2011,rechtsman_prl_2013},
{ similar band }structures can also be obtained by the tight-binding method.
In addition, the full Floquet band structure and the edge band structure can also be analyzed {using the} tight-binding method \cite{takashi_prb_2009,fang_oe_2013}.

It is clear that the induced RI modulation exhibits a honeycomb profile,
and the PBG structure contains 6 Dirac points and cones at the corners of the first Brillouin zone.
However, different from the previous systems considered, the RI change, as well as the PBG structures,
can be easily adjusted by the frequency detunings ($\Delta_1$ and $\Delta_2$),
which is one of the main advantages of the current system.
Figures \ref{fig2}(a1)-\ref{fig2}(a5) and \ref{fig2}(b1)-\ref{fig2}(b5) exhibit RIs,
as well as the corresponding PBG structures, with different frequency detunings.
In Fig. \ref{fig2}(a1), the RI at the honeycomb lattice sites is the smallest
{(also in the case of Fig. \ref{fig2}(a2))},
and there are Dirac cones in the corresponding PBG, as shown in Fig. \ref{fig2}(b1).
If $\Delta_1=\Delta_2=0 $, the linear susceptibility in Eq. (\ref{eq2}) is imaginary,
so that the RI is always 1 {in} the transverse plane, as shown in Fig. \ref{fig2}(a3).
Since there is no RI change, {the honeycomb lattice disappears and}
there are no Dirac cones in the PBG, as displayed in Fig. \ref{fig2}(b3),
in which the edges of the bands merge with each other.
Note that for other cases with $\Delta_1-\Delta_2=0 $, the RI will not be 1 everywhere.
If $\Delta_1<0 $, the RI change at the honeycomb lattice sites will be the smallest,
while if $\Delta_1>0 $, {it will be} the biggest.
Therefore, even if the first two bands seemingly merge with each other,
there are still Dirac cones, which is different from the case shown in Fig. \ref{fig2}(b3).

If ${\Delta_1=-\Delta_2=5} $ MHz, the RI change at the honeycomb lattice sites is the biggest, as exhibited in Fig. \ref{fig2}(a4),
and the corresponding PBG in Fig. \ref{fig2}(b4) shows a big band gap between the first two bands,
so that the Dirac cones disappear.
If we adjust the frequency detunings to ${\Delta_1=-\Delta_2=10} $ MHz, as shown in Figs. \ref{fig2}(a5) and \ref{fig2}(b5),
the RI change is still the biggest at the lattice sites,
however the first two bands become almost degenerate.
If we display the first two bands exclusively, as shown in the inset in Fig. \ref{fig2}(b5),
we find there are still 6 Dirac cones.
However, the width of the first two bands is too small to be resolved in comparison with the big band gap.

It is worthwhile explaining why the frequency detunings can determine whether the honeycomb RI PBG structure has Dirac cones or not.
In Figs. \ref{fig2}(a1) and \ref{fig2}(a2), the effective potential or the RI change at the lattice sites is deep enough to ensure the appearance of Dirac cones.
With increasing frequency detunings, the potential wells at or around lattice sites become smaller,
which cannot support the Dirac cones anymore (Figs. \ref{fig2}(a3) and \ref{fig2}(a4)).
Further increasing the frequency detunings will produce potential barriers at lattice sites,
so that Dirac cones can appear again in the PBG structure (Fig. \ref{fig2}(a5)).
According to Eq. (\ref{eq4}), $V_0$ is related to the potential depth, so for the same frequency detunings,
the bigger $V_0$ will make the observation of Dirac cones easier.

\section{Photonic Floquet topological insulator}\label{PFTI}

As demonstrated previously, honeycomb lattices possess edge states when they have finite size
and exhibit Dirac cones in their PBG structures \cite{rechtsman_nature_2013, rechtsman_prl_2013, plotnik_nm_2014}.
If we transform the coordinates by
$x'= x+R\cos(\omega z)$,
$y'= y+R\sin(\omega z)$, and
$z'= z$,
where $R$ is the helix radius and $\omega$ the frequency of rotation,
the lattice sites of the interference honeycomb pattern will be spiraling along the $z$ direction.
According to the discussion on the relation between Floquet modes and helical waveguides in Ref. \cite{rechtsman_nature_2013},
{such a spiraling honeycomb lattice formed in the three-beam interference can serve as a PFTI generated in a multi-level atomic ensemble.}

{Generally, there are three methods that can be utilized for the creation of a helical waveguide system:
(i) femtosecond laser writing technique;
(ii) holographic lithography method;
and (iii) nonlinear phase shift (NPS) modulation.
The first method is only valid for solid materials \cite{rechtsman_nature_2013}.
In the second method, $6+1$ beams with different polarizations are used \cite{kennedy_nanolett_2002,pang_oe_2005,seet_am_2005},
but this quite complicated method is also mostly developed for solid materials.
Thus far, helical waveguides in atomic ensembles have not been discussed.
Hence, we propose the third feasible method for realization of a helical honeycomb pattern -- the NPS modulation \cite{agrawal_prl_1990,hickmann_prl_1992},
achieved by employing an additional controlling field in an atomic system, as shown in Fig. \ref{fig1}(a).}

The added controlling field $E_3 $ will split $|+\rangle $ into ${|+\pm\rangle }$ with eigenfrequencies
$ \ell_{|+\pm\rangle} = (-\Delta_2/2 + \sqrt{\Delta_2{}^2/4+|G|^2}) + (\Delta'_3/2 \pm \sqrt{\Delta'_3{}^2/4+|G_3|^2})  $,
in which $\Delta_3' = \Delta_3 + ( \Delta_2 /2 - \sqrt{\Delta_2{}^2/4 + |G|^2}  ) $;
$\Delta_3$ is the frequency detuning of $E_3$ and $G_3$ is the Rabi frequency of $E_3$.
The NPS can be written as $ S_{NL}(r,\,\phi,\,z) = 2 k_2 n_2^X I_{3} e^{-[r^2+l^2+2rl\cos(\phi-\phi')]} z/n_0$
in the cylindrical coordinates, with ${r=\sqrt{x^2+y^2} }$ and $\phi=\arctan(y/x) $,
where $l$ and $\phi'$ are the distance and angle of the controlling field relative to the lattice site.
Here $n_2^X ={\rm Re} \{\chi^{(3)X} \}/(\epsilon_0 cn_0)$ is the cross-Kerr nonlinear index from the controlling field,
$\chi^{(3)X} = N \wp_{10}^2 \wp_{13}^2 \rho^{(3)}_{10} /(\hbar^3 \epsilon_0 G_p G_3^2) $,
$\rho^{(3)}_{10} $ is the corresponding density-matrix element, and $I_3$ the intensity of the controlling field.

\begin{figure}[htbp]
\centering
  \includegraphics[width=0.3\columnwidth]{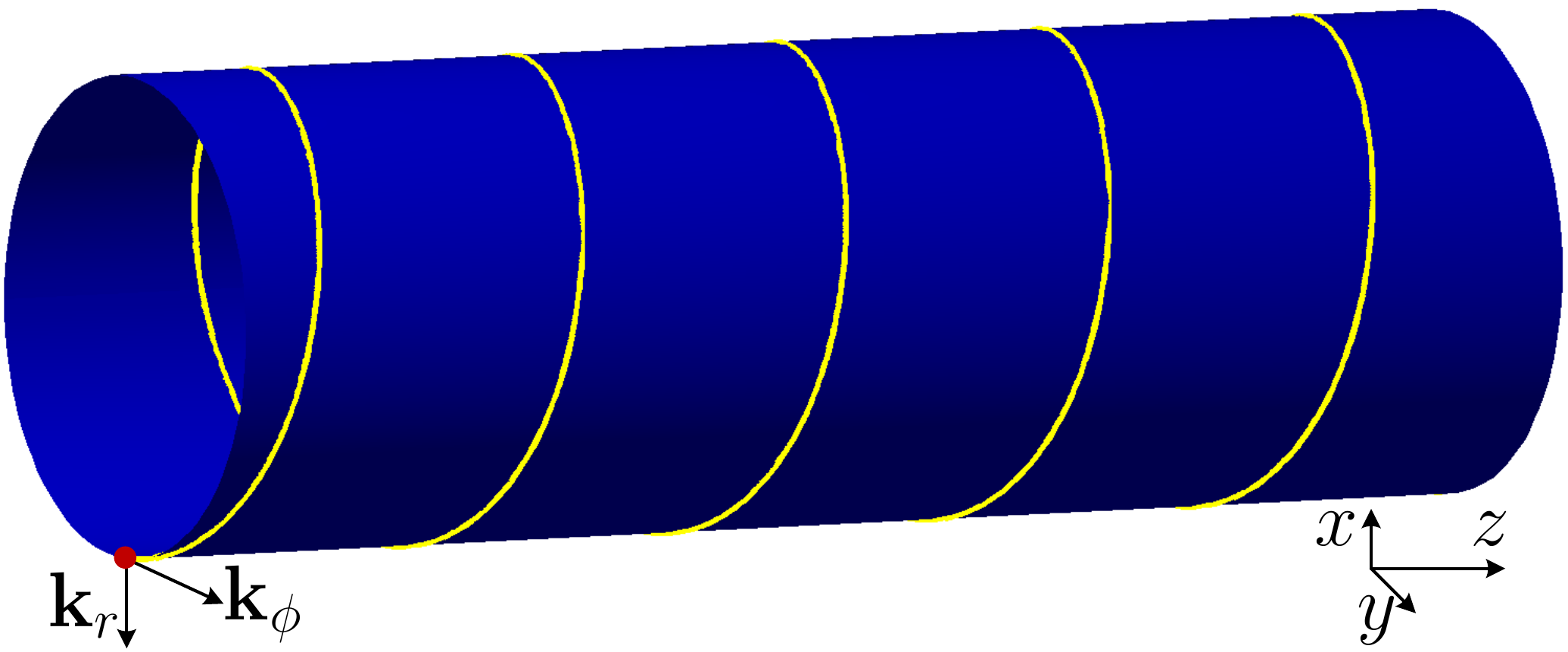}
  \caption{{ (Color online)
  Illustrating the formation of a helical waveguide due to NPS modulation.
  Each lattice site forms one waveguide.
  }}
  \label{helical}
\end{figure}

The NPS will introduce a transverse wave vector $\delta {\bf k}_\bot  (r,\,\phi) = \hat r (\partial S_{NL} / \partial r) +
\hat \phi (\partial S_{NL} / \partial \phi)/r = {\bf k}_r + {\bf k}_\phi$, with $\hat r$ and $\hat\phi$ being the unit vectors.
Specifically, the momenta can be written as ${\bf k}_r= - 2 S_{NL}[r + l\cos (\phi-\phi' )]\hat r$ and ${\bf k}_\phi=-2 S_{NL} l \sin(\phi-\phi' ) \hat \phi$,
which will determine the radial and azimuthal movements of the lattice sites, respectively.
Therefore, if the NPS is imposed on each honeycomb lattice site, the site will be driven to rotate in the transverse plane and spiral along the propagation direction $z$, { as elucidated by the curved waveguide in Fig. \ref{helical}.}
Furthermore, the spiraling direction (clockwise or anticlockwise),
the period of the spiral and the radius of the helical pattern can be all adjusted through controlling the beam intensity $I_3$
and the nonlinear index $n_2^X $.
In a hot atomic vapor, the propagation distance $z$ is effectively related to the atomic density,
which can be easily controlled by the temperature \cite{zhang_prl_2011, wu_pra_2013}.
{There, one can use a beam shaper to prepare
the coupling fields with certain profiles. With the spatially
shaped beams, the interference of the coupling fields will
lead to a honeycomb lattice with well-defined boundaries.
On the other hand, for cold atomic media, the boundary can
be obtained by using the boundary of the magneto-optical
trap, which is wide in comparison to the wavelength of light.}
For cold cigar-shaped atomic clouds, one can adjust the length of the sample with the trap potential.

\begin{figure*}[htbp]
\centering
  \includegraphics[width=0.7\textwidth]{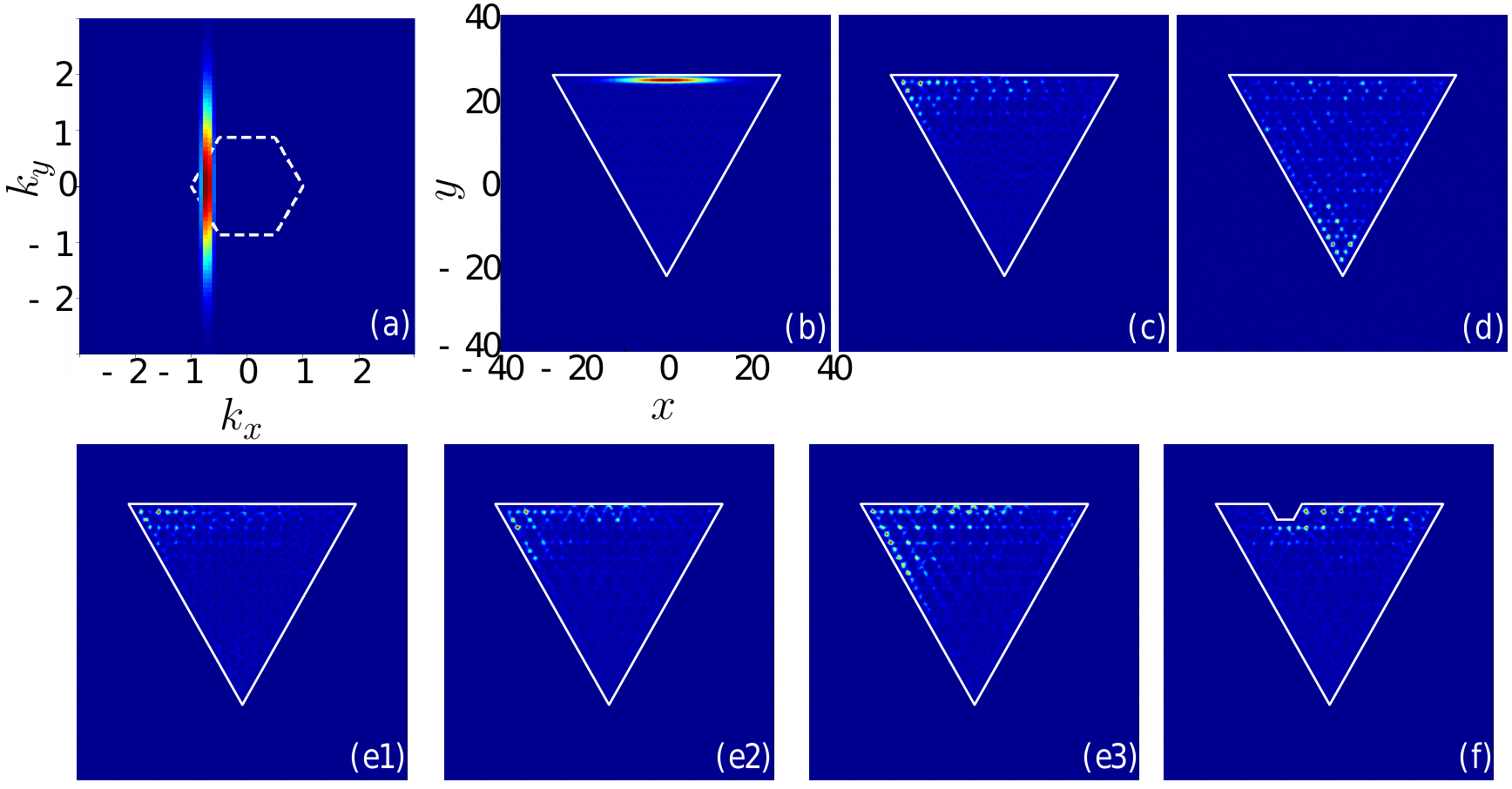}
  \caption{(Color online) (a) Input beam exhibited in Fourier space. The dashed hexagon is the first Brillouin zone.
  (b)-(d) Simulated probe beam intensity distributions when propagated to distances $z=0$, $z\approx6.3~\mu \rm m$, and $z\approx18.6~\mu \rm m$, respectively.
  The inverted triangle presents the PFTI with modulated RI and zigzag boundaries.
  The parameters are ${\Delta_1=-\Delta_2=-10}$ MHz, $V_0=150$, $R \approx 24.8 ~\rm nm $, and $\omega/(2\pi) \approx 0.8~ \rm GHz$ (the period is $\sim1.2$ nm).
  (e1)-(e3) Same as (c) but under ${\Delta_1=-\Delta_2=-11}$ MHz,
  ${\Delta_1=-\Delta_2=-9}$ MHz, and ${\Delta_1=-\Delta_2=-8}$ MHz, respectively.
  (f) Beam propagates to $z\approx 4.7~\mu \rm m$ with a disorder displayed in the PFTI.}
  \label{fig3}
\end{figure*}

If one launches a probe beam into the medium along one edge \cite{plotnik_nm_2014},
the corresponding optical properties can be easily studied by propagating the beam according to Eq. (\ref{eq4}).
{Because of} the distribution of zigzag edge states in momentum space,
one should launch the incident beam obliquely.
In Fig. \ref{fig3}(a), we show the obliquely incident beam in momentum space with the dashed hexagon being the first Brillouin zone.
It is clear that the obliquely incident beam may excite the zigzag edge state.
In real space, the incident beam is shown in Fig. \ref{fig3}(b),
in which the inverted triangle is the constructed PFTI.
When the probe beam propagates to a distance of $z\approx6.3~\mu \rm m$, the intensity distribution is numerically stimulated and exhibited in Fig. \ref{fig3}(c).
Comparing Fig. \ref{fig3}(c) with Fig. \ref{fig3}(b),
one can see that the beam moves to the left (counterclockwise) along the zigzag edge without scattering energy into the bulk of PFTI.
When the beam further propagates to $z\approx18.6~\mu \rm m$,
it moves to the bottom corner of the PFTI (Fig. \ref{fig3}(d)),
and still there is nearly no energy scattered into PFTI.
The phenomenon displayed in Figs. \ref{fig3}(b)-\ref{fig3}(d) can be naturally explained by the following two {arguments}:
(I) The zigzag edge state is excited by the obliquely incident beam,
so that the confinement of the edge state prohibits the scattering of the beam {into the bulk} during propagation.
(II) The honeycomb lattice sites are spiraling along the $z$ direction;
such a structure helps the edge state attain an effective velocity along the propagation distance $z$,
which enables the probe beam to move one-way (anticlockwise) along the zigzag edge.

\section{Discussion}\label{discussion}

\begin{figure*}[htbp]
\centering
  \includegraphics[width=0.75\textwidth]{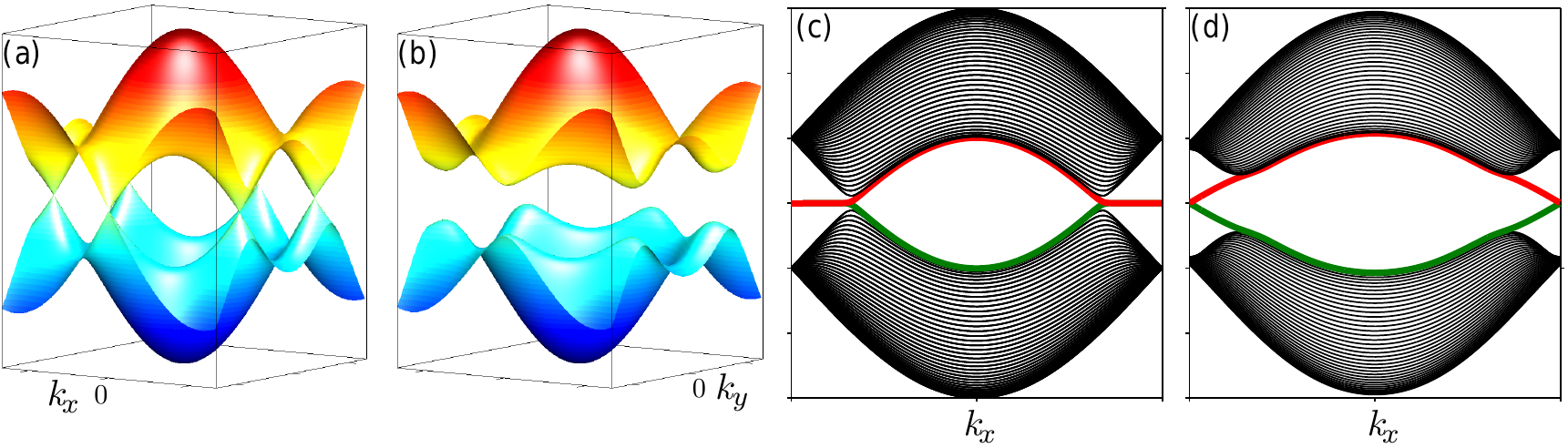}
  \caption{{ (Color online)
  (a) Band structure when the honeycomb waveguides are not spiraling along the propagation direction.
  (b) Same as (a) but the waveguides are helical along the propagation direction.
  (c) Corresponding to (a); edge band structure of the strained honeycomb waveguides with zigzag boundaries.
  (d) Same as (c), but corresponding to (b).
  }}
  \label{edge}
\end{figure*}

{
In Fig. \ref{edge}, we show the full band structure and the corresponding edge band structure for a spiraling and non-spiraling lattice.
One can see that Dirac points in Fig. \ref{edge}(a) are eliminated in Fig. \ref{edge}(b), creating a PFTI.
The topological invariant for each band shown in Fig. \ref{edge}
can be evaluated by the Chern number \cite{haldane_prl_2008,wang_nature_2009,zak_prl_1989},
which is given by
\[
C = \frac{1}{2\pi} \int_{\rm BZ} d^2 k \nabla_{\bf k} \times {\mathcal A ({\bf k})},
\]
where the $k$-space integral is performed over the first Brillouin zone and the Berry connection is given by
\[
{\mathcal A} ({\bf k}) = i \langle u_{\bf k}({\bf r}) | \nabla_{\bf k} | u_{\bf k}({\bf r}) \rangle = i \int d^2r u_{\bf k}^*({\bf r}) \cdot [\nabla_{\bf k} u_{\bf k}({\bf r}) ],
\]
with $u_{\bf k}({\bf r}) $ being the periodic part of the Bloch function.
Since the full Floquet band structure [Figs. \ref{edge}(a) and \ref{edge}(b)],
the edge band structure [curves in the band gap in Figs. \ref{edge}(c) and \ref{edge}(d)],
and the analysis of }topological invariants are quite similar to those given in Ref. \cite{rechtsman_nature_2013},
they will not be further explored here.
Even though the corresponding topological invariant (or the topological protection) is demonstrated by an integer Chern number,
we believe that such topological invariants can be well elucidated by looking at Fig. \ref{fig3}.

When taking $\Delta_1=-\Delta_2=-11 $ MHz (Fig. \ref{fig3}(e1)), $\Delta_1=-\Delta_2=-9 $ MHz (Fig. \ref{fig3}(e2)),
and $\Delta_1=-\Delta_2=-8 $ MHz (Fig. \ref{fig3}(e3)), respectively,
we find that the beams in Figs. \ref{fig3}(e2) and \ref{fig3}(e3) move faster along the zigzag edge than the cases displayed in Figs. \ref{fig3}(c) and \ref{fig3}(e1),
i.e., the velocity is bigger if the absolute values of the frequency detunings are smaller.
The reason is quite clear -- smaller frequency detunings will lead to a shallower potential (the dips in the RI are smaller),
so that the beam can overpass successive potential  barriers easier.
In Figs. \ref{fig2}(a5) and \ref{fig2}(b5), the RI change also displays Dirac cones in the corresponding momentum space.
However, we could not construct a PFTI using the parameters from Figs. \ref{fig2}(a5) and \ref{fig2}(b5).
Comparing Fig. \ref{fig2}(a1) with Fig. \ref{fig2}(a5) and considering the beam localization in Fig. \ref{fig3},
one can see that the beam is trapped in the regions around the dips in RI,
which present potential wells.
In Fig. \ref{fig2}(a5), the  potential barriers are discrete peaks,
which are rather hard for the beam to overcome and obtain transverse velocity.
Note that the positive change in RI, $\Delta n(x,\,y,\,z)>0$, (the red regions in Fig. \ref{fig2}) indicates the attractive of focusing effective potential, whereas
$\Delta n(x,\,y,\,z)<0$ (the blue regions) indicates repulsive or defocusing potential.

On the other hand, when the first two bands are almost degenerate, the edge states will be too weak to confine the beam to the edge.
This tendency is already visible in Fig. \ref{fig3}(e3), where the detunings are smaller and the
scattering into bulk is larger.
Various numerical simulations indicate that ${\Delta_1=-\Delta_2>0} $ would not support the formation of PFTI.
So, the appearance of Dirac cones in momentum space is not the sufficient condition for realizing PFTI.
By fixing $\Delta_1=-\Delta_2=-10 $ MHz, we can change the intensities of the coupling fields to investigate the properties of the constructed PFTI.
We find that the velocity of the moving beam increases with the increasing intensities of the coupling fields.
If the PFTI possesses a disorder, as shown in Fig. \ref{fig3}(f), the beam will move around it, due to the topological protection.

\begin{figure*}[htbp]
\centering
  \includegraphics[width=0.7\textwidth]{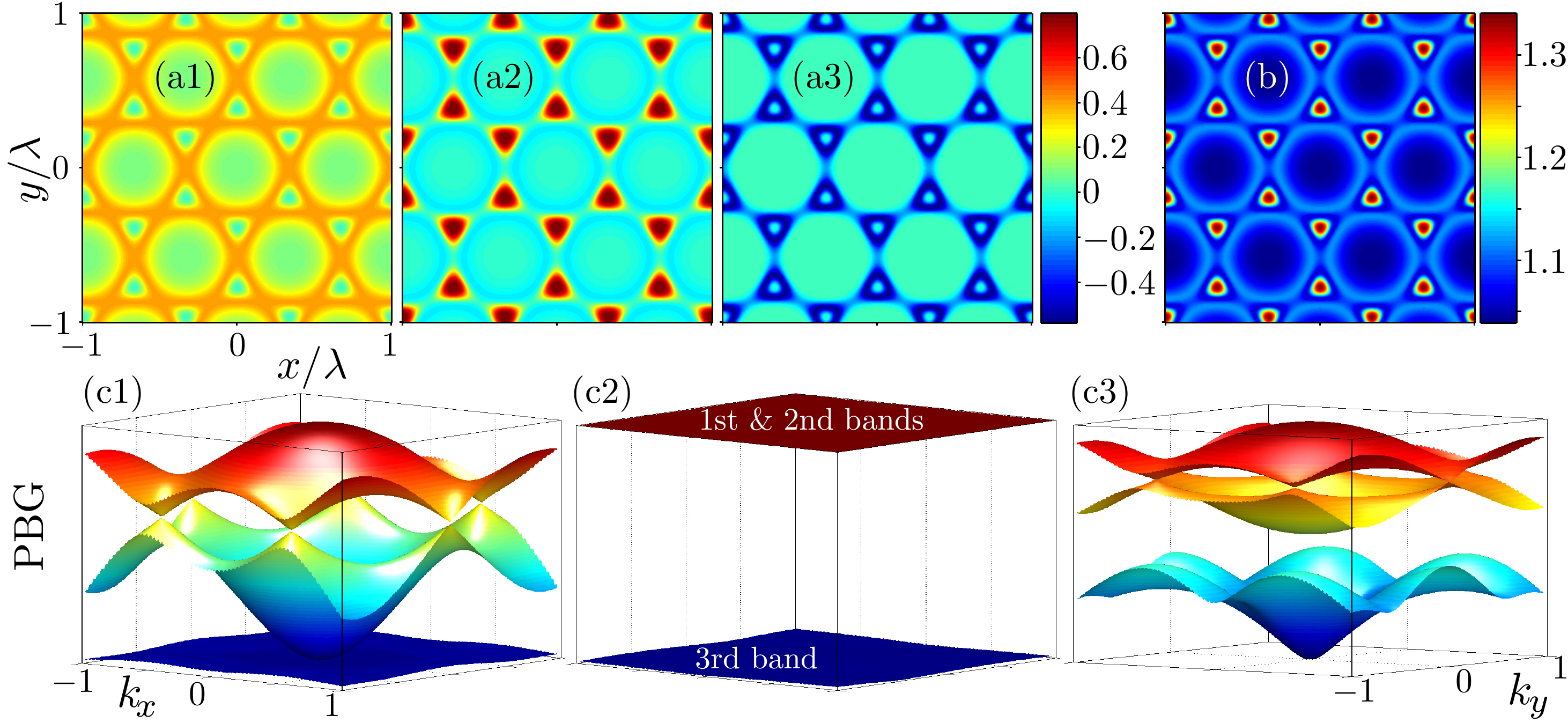}
  \caption{(Color online) (a1)-(a3) Real parts of $\chi_p^{(1)} $, $\chi_p^{(3)} |E_2|^2 $ and $\chi_p^{(5)} |E_2|^4 $, respectively.
  (b) Modulated RI patterns with $\chi_p^{(1)} $, $\chi_p^{(3)} |E_2|^2 $ and $\chi_p^{(5)} |E_2|^4 $ all considered. Parameters are $\Delta_1=0 $, $\Delta_2=10 $ MHz and $G_2=10 $ MHz.
  (c) PBG structures for RI with $\chi_p^{(1)}$ (c1), $\chi_p^{(1)} + \chi_p^{(3)} |E_2|^2 $ (c2), and $\chi_p^{(1)} + \chi_p^{(3)} |E_2|^2 + \chi_p^{(5)} |E_2|^4 $ (c3) considered, respectively.
  }
  \label{fig4}
\end{figure*}

When the coupling beam intensities are high enough,
the third- and fifth-order nonlinear susceptibilities $({\chi_p^{(3)}|E_2|^2}$ and ${\chi_p^{(5)}|E_2|^4})$ come into play and
should be taken into account \cite{mihalache_pre_2003,desyatnikov_pre_2005, michinel_prl_2006, wu_pra_2013, paredes_prl_2014}.
Then, the total susceptibility should be modified to
\[
\begin{split}
\Delta n  (x,\,y) \approx  \delta n_0 +\\
\delta n_1 [2 \cos(3k_2 x/2) \cos(\sqrt{3} k_2 y/2) + \cos(\sqrt{3} k_2 y)] +   \\
\delta n_2 [2 \cos(3k_2 x/2) \cos(\sqrt{3} k_2 y/2) + \cos(\sqrt{3} k_2 y)]^2 + \\
\delta n_4 [{2 \cos(3k_2 x/2) \cos(\sqrt{3} k_2 y/2) + \cos(\sqrt{3} k_2 y)}]^3,
\end{split}
\]
with $\delta n_2$ and $\delta n_4$ being the higher-order nonlinearity coefficients,
which are connected with the Rabi frequencies and frequency detunings.
Here, we have \[\delta n_0 \approx \sqrt{1 + {\rm Re} \{(1 - \tau + \tau^2) \eta \} }, \]
\[\delta n_1 \approx -{\rm Re} \{ (1 - 2\tau + 3\tau^2) \eta \xi \} / (2\delta n_0), \]
\[\delta n_2 \approx {\rm Re} \{ (1 - 3\tau + 6\tau^2) \eta \xi^2 \} / (2\delta n_0), \] and
\[\delta n_3 \approx -{\rm Re} \{ (1 - 4\tau + 10\tau^2) \eta \xi^3 \} / (2\delta n_0), \]
with $\tau=G_2^2/(d_{10}d_{20} + G_2^2) $.
With an increasing $G_2$, the influence of higher-order nonlinear susceptibilities grows,
which will modify the RI patterns significantly.
Taking ${\Delta_1=-\Delta_2=-10}$ MHz as an example, $G_2>19.5$ MHz will make the RI complex,
so that the beam may undergo gain or loss during propagation \cite{szameit_pra_2011},
which may provide a way to study the $\mathcal{PT}$ symmetry in atomic ensembles \cite{hang_prl_2013, sheng_pra_2013}.

In Figs. \ref{fig4}(a1)-\ref{fig4}(a3),
we plot the real parts of the susceptibilities $\chi_p^{(1)} $, $\chi_p^{(3)} |E_2|^2 $ and $\chi_p^{(5)} |E_2|^4 $, respectively,
under the same color scale, with $\Delta_1=0 $, $\Delta_2=10 $ MHz and $G_2=10 $ MHz.
It is clear that the signs of the first- and fifth-order susceptibilities are the same,
while the third-order susceptibility has the opposite sign.
The RI values of $\chi_p^{(1)} $ at the lattice sites will change when the modifications of $\chi_p^{(3)} |E_2|^2 $ and $\chi_p^{(5)} |E_2|^4 $ are {added}.
In Fig. \ref{fig4}(b), we display the total RI pattern with all three susceptibilities considered.
Thus, the RI values at the lattice sites can be controlled through manipulating the frequency detunings and intensities of the coupling fields.

The PBG structures of the RI for $\chi_p^{(1)} $, $\chi_p^{(1)} + \chi_p^{(3)} |E_2|^2$ and
$\chi_p^{(1)} + \chi_p^{(3)} |E_2|^2 + \chi_p^{(5)} |E_2|^4 $ are shown in Figs. \ref{fig4}(c1),  \ref{fig4}(c2) and  \ref{fig4}(c3), respectively.
Dirac cones clearly appear in Fig.  \ref{fig4}(c1).
However, in Fig.  \ref{fig4}(c2) the first two bands are nearly degenerate and flat, with a wide band gap between them and the third band \cite{sun_prl_2011},
which is quite similar to the case in Fig. \ref{fig2}(b5).
The flat bands mean that only the light with very special propagation constants can be allowed to couple among {the sites of honeycomb lattice}.
The reason for the appearance of flat bands is that the third-order nonlinear susceptibility makes the bands collapse \cite{crespi_njp_2013}.
As a result, the PFTI cannot be formed when the third-order nonlinear susceptibility is included {with given} parameters.
However, when the fifth-order nonlinear susceptibility is also included,
{one sees} that the Dirac cones reappear again in the first two bands, as shown in Fig. \ref{fig4}(c3).
Therefore, the third-order nonlinear susceptibility shuts down and the fifth-order nonlinear susceptibility reopens the Dirac cones.
Therefore, the high-order nonlinear susceptibilities can serve as a kind of switch,
which determines the appearance {and disappearance} of Dirac cones in the momentum space,
and so the formation of PFTIs in the system.

\section{Conclusion}\label{conclude}

In summary, we have proposed a scheme for construction of PFTIs in multi-level atomic vapor ensembles.
The formed PFTIs in atomic ensembles can be easily controlled and reconfigured by adjusting the frequency detunings,
coupling field intensities, and high-order nonlinear susceptibilities,
which shows the advantages of using atomic ensembles to study PFTI properties in comparison with other solid media.
The PFTIs should also exist in other types of multi-level atomic systems.
{Easy controllability of the spiraling and the
switching property of higher-order susceptibilities establish
a new platform for better understanding of the topological
protection and open potential photonic device applications
of PFTIs.}

\section*{acknowledgement}
This work was supported by the 973 Program (2012CB921804),
KSTIT of Shaanxi province (2014KCT-10),
CPSF (2014T70923, 2012M521773),
NSFC (61308015, 11474228),
NSFC of Shaanxi province (2014JQ8341),
and the NPRP 6-021-1-005 project of the Qatar National Research Fund
(a member of the Qatar Foundation).
Yiqi Zhang appreciates the helpful discussions with Dr. Ruimin Wang.


\begin{thebibliography}{50}%
\makeatletter
\providecommand \@ifxundefined [1]{%
 \@ifx{#1\undefined}
}%
\providecommand \@ifnum [1]{%
 \ifnum #1\expandafter \@firstoftwo
 \else \expandafter \@secondoftwo
 \fi
}%
\providecommand \@ifx [1]{%
 \ifx #1\expandafter \@firstoftwo
 \else \expandafter \@secondoftwo
 \fi
}%
\providecommand \natexlab [1]{#1}%
\providecommand \enquote  [1]{``#1''}%
\providecommand \bibnamefont  [1]{#1}%
\providecommand \bibfnamefont [1]{#1}%
\providecommand \citenamefont [1]{#1}%
\providecommand \href@noop [0]{\@secondoftwo}%
\providecommand \href [0]{\begingroup \@sanitize@url \@href}%
\providecommand \@href[1]{\@@startlink{#1}\@@href}%
\providecommand \@@href[1]{\endgroup#1\@@endlink}%
\providecommand \@sanitize@url [0]{\catcode `\\12\catcode `\$12\catcode
  `\&12\catcode `\#12\catcode `\^12\catcode `\_12\catcode `\%12\relax}%
\providecommand \@@startlink[1]{}%
\providecommand \@@endlink[0]{}%
\providecommand \url  [0]{\begingroup\@sanitize@url \@url }%
\providecommand \@url [1]{\endgroup\@href {#1}{\urlprefix }}%
\providecommand \urlprefix  [0]{URL }%
\providecommand \Eprint [0]{\href }%
\providecommand \doibase [0]{http://dx.doi.org/}%
\providecommand \selectlanguage [0]{\@gobble}%
\providecommand \bibinfo  [0]{\@secondoftwo}%
\providecommand \bibfield  [0]{\@secondoftwo}%
\providecommand \translation [1]{[#1]}%
\providecommand \BibitemOpen [0]{}%
\providecommand \bibitemStop [0]{}%
\providecommand \bibitemNoStop [0]{.\EOS\space}%
\providecommand \EOS [0]{\spacefactor3000\relax}%
\providecommand \BibitemShut  [1]{\csname bibitem#1\endcsname}%
\let\auto@bib@innerbib\@empty
\bibitem [{\citenamefont {Kane}\ and\ \citenamefont
  {Mele}(2005)}]{kane_prl_2005}%
  \BibitemOpen
  \bibfield  {author} {\bibinfo {author} {\bibfnamefont {C.~L.}\ \bibnamefont
  {Kane}}\ and\ \bibinfo {author} {\bibfnamefont {E.~J.}\ \bibnamefont
  {Mele}},\ }\href {\doibase 10.1103/PhysRevLett.95.226801} {\bibfield
  {journal} {\bibinfo  {journal} {Phys. Rev. Lett.}\ }\textbf {\bibinfo
  {volume} {95}},\ \bibinfo {pages} {226801} (\bibinfo {year}
  {2005})}\BibitemShut {NoStop}%
\bibitem [{\citenamefont {Hsieh}\ \emph {et~al.}(2008)\citenamefont {Hsieh},
  \citenamefont {Qian}, \citenamefont {Wray}, \citenamefont {Xia},
  \citenamefont {Hor}, \citenamefont {Cava},\ and\ \citenamefont
  {Hasan}}]{hsieh_nature_2008}%
  \BibitemOpen
  \bibfield  {author} {\bibinfo {author} {\bibfnamefont {D.}~\bibnamefont
  {Hsieh}}, \bibinfo {author} {\bibfnamefont {D.}~\bibnamefont {Qian}},
  \bibinfo {author} {\bibfnamefont {L.}~\bibnamefont {Wray}}, \bibinfo {author}
  {\bibfnamefont {Y.}~\bibnamefont {Xia}}, \bibinfo {author} {\bibfnamefont
  {Y.~S.}\ \bibnamefont {Hor}}, \bibinfo {author} {\bibfnamefont {R.~J.}\
  \bibnamefont {Cava}}, \ and\ \bibinfo {author} {\bibfnamefont {M.~Z.}\
  \bibnamefont {Hasan}},\ }\href {\doibase 10.1038/nature06843} {\bibfield
  {journal} {\bibinfo  {journal} {Nature}\ }\textbf {\bibinfo {volume} {452}},\
  \bibinfo {pages} {970} (\bibinfo {year} {2008})}\BibitemShut {NoStop}%
\bibitem [{\citenamefont {Hasan}\ and\ \citenamefont
  {Kane}(2010)}]{hasan_rmp_2010}%
  \BibitemOpen
  \bibfield  {author} {\bibinfo {author} {\bibfnamefont {M.~Z.}\ \bibnamefont
  {Hasan}}\ and\ \bibinfo {author} {\bibfnamefont {C.~L.}\ \bibnamefont
  {Kane}},\ }\href {\doibase 10.1103/RevModPhys.82.3045} {\bibfield  {journal}
  {\bibinfo  {journal} {Rev. Mod. Phys.}\ }\textbf {\bibinfo {volume} {82}},\
  \bibinfo {pages} {3045} (\bibinfo {year} {2010})}\BibitemShut {NoStop}%
\bibitem [{\citenamefont {Qi}\ and\ \citenamefont {Zhang}(2011)}]{qi_rmp_2011}%
  \BibitemOpen
  \bibfield  {author} {\bibinfo {author} {\bibfnamefont {X.-L.}\ \bibnamefont
  {Qi}}\ and\ \bibinfo {author} {\bibfnamefont {S.-C.}\ \bibnamefont {Zhang}},\
  }\href {\doibase 10.1103/RevModPhys.83.1057} {\bibfield  {journal} {\bibinfo
  {journal} {Rev. Mod. Phys.}\ }\textbf {\bibinfo {volume} {83}},\ \bibinfo
  {pages} {1057} (\bibinfo {year} {2011})}\BibitemShut {NoStop}%
\bibitem [{\citenamefont {Liu}\ \emph {et~al.}(2011)\citenamefont {Liu},
  \citenamefont {Yin}, \citenamefont {Ulin-Avila}, \citenamefont {Geng},
  \citenamefont {Zentgraf}, \citenamefont {Ju}, \citenamefont {Wang},\ and\
  \citenamefont {Zhang}}]{liu_nature_2011}%
  \BibitemOpen
  \bibfield  {author} {\bibinfo {author} {\bibfnamefont {M.}~\bibnamefont
  {Liu}}, \bibinfo {author} {\bibfnamefont {X.}~\bibnamefont {Yin}}, \bibinfo
  {author} {\bibfnamefont {E.}~\bibnamefont {Ulin-Avila}}, \bibinfo {author}
  {\bibfnamefont {B.}~\bibnamefont {Geng}}, \bibinfo {author} {\bibfnamefont
  {T.}~\bibnamefont {Zentgraf}}, \bibinfo {author} {\bibfnamefont
  {L.}~\bibnamefont {Ju}}, \bibinfo {author} {\bibfnamefont {F.}~\bibnamefont
  {Wang}}, \ and\ \bibinfo {author} {\bibfnamefont {X.}~\bibnamefont {Zhang}},\
  }\href {\doibase 10.1038/nature10067} {\bibfield  {journal} {\bibinfo
  {journal} {Nature}\ }\textbf {\bibinfo {volume} {474}},\ \bibinfo {pages}
  {64} (\bibinfo {year} {2011})}\BibitemShut {NoStop}%
\bibitem [{\citenamefont {Yu}\ \emph {et~al.}(2013)\citenamefont {Yu},
  \citenamefont {Zhang}, \citenamefont {Wang}, \citenamefont {Zhao},
  \citenamefont {Wang}, \citenamefont {Wen}, \citenamefont {Zhang},\ and\
  \citenamefont {Wang}}]{yu_lpor_2013}%
  \BibitemOpen
  \bibfield  {author} {\bibinfo {author} {\bibfnamefont {H.}~\bibnamefont
  {Yu}}, \bibinfo {author} {\bibfnamefont {H.}~\bibnamefont {Zhang}}, \bibinfo
  {author} {\bibfnamefont {Y.}~\bibnamefont {Wang}}, \bibinfo {author}
  {\bibfnamefont {C.}~\bibnamefont {Zhao}}, \bibinfo {author} {\bibfnamefont
  {B.}~\bibnamefont {Wang}}, \bibinfo {author} {\bibfnamefont {S.}~\bibnamefont
  {Wen}}, \bibinfo {author} {\bibfnamefont {H.}~\bibnamefont {Zhang}}, \ and\
  \bibinfo {author} {\bibfnamefont {J.}~\bibnamefont {Wang}},\ }\href {\doibase
  10.1002/lpor.201300084} {\bibfield  {journal} {\bibinfo  {journal} {Laser
  Photon. Rev.}\ }\textbf {\bibinfo {volume} {7}},\ \bibinfo {pages} {L77}
  (\bibinfo {year} {2013})}\BibitemShut {NoStop}%
\bibitem [{\citenamefont {Liang}\ and\ \citenamefont
  {Chong}(2013)}]{liang_prl_2013}%
  \BibitemOpen
  \bibfield  {author} {\bibinfo {author} {\bibfnamefont {G.~Q.}\ \bibnamefont
  {Liang}}\ and\ \bibinfo {author} {\bibfnamefont {Y.~D.}\ \bibnamefont
  {Chong}},\ }\href {\doibase 10.1103/PhysRevLett.110.203904} {\bibfield
  {journal} {\bibinfo  {journal} {Phys. Rev. Lett.}\ }\textbf {\bibinfo
  {volume} {110}},\ \bibinfo {pages} {203904} (\bibinfo {year}
  {2013})}\BibitemShut {NoStop}%
\bibitem [{\citenamefont {Khanikaev}\ \emph {et~al.}(2012)\citenamefont
  {Khanikaev}, \citenamefont {Mousavi}, \citenamefont {Tse}, \citenamefont
  {Kargarian}, \citenamefont {MacDonald},\ and\ \citenamefont
  {Shvets}}]{khanikaev_nm_2012}%
  \BibitemOpen
  \bibfield  {author} {\bibinfo {author} {\bibfnamefont {A.~B.}\ \bibnamefont
  {Khanikaev}}, \bibinfo {author} {\bibfnamefont {S.~H.}\ \bibnamefont
  {Mousavi}}, \bibinfo {author} {\bibfnamefont {W.-K.}\ \bibnamefont {Tse}},
  \bibinfo {author} {\bibfnamefont {M.}~\bibnamefont {Kargarian}}, \bibinfo
  {author} {\bibfnamefont {A.~H.}\ \bibnamefont {MacDonald}}, \ and\ \bibinfo
  {author} {\bibfnamefont {G.}~\bibnamefont {Shvets}},\ }\href {\doibase
  10.1038/nmat3520} {\bibfield  {journal} {\bibinfo  {journal} {Nat. Mater.}\
  }\textbf {\bibinfo {volume} {12}},\ \bibinfo {pages} {233} (\bibinfo {year}
  {2012})}\BibitemShut {NoStop}%
\bibitem [{\citenamefont {Rechtsman}\ \emph
  {et~al.}(2013{\natexlab{a}})\citenamefont {Rechtsman}, \citenamefont
  {Zeuner}, \citenamefont {Plotnik}, \citenamefont {Lumer}, \citenamefont
  {Podolsky}, \citenamefont {Dreisow}, \citenamefont {Nolte}, \citenamefont
  {Segev},\ and\ \citenamefont {Szameit}}]{rechtsman_nature_2013}%
  \BibitemOpen
  \bibfield  {author} {\bibinfo {author} {\bibfnamefont {M.~C.}\ \bibnamefont
  {Rechtsman}}, \bibinfo {author} {\bibfnamefont {J.~M.}\ \bibnamefont
  {Zeuner}}, \bibinfo {author} {\bibfnamefont {Y.}~\bibnamefont {Plotnik}},
  \bibinfo {author} {\bibfnamefont {Y.}~\bibnamefont {Lumer}}, \bibinfo
  {author} {\bibfnamefont {D.}~\bibnamefont {Podolsky}}, \bibinfo {author}
  {\bibfnamefont {F.}~\bibnamefont {Dreisow}}, \bibinfo {author} {\bibfnamefont
  {S.}~\bibnamefont {Nolte}}, \bibinfo {author} {\bibfnamefont
  {M.}~\bibnamefont {Segev}}, \ and\ \bibinfo {author} {\bibfnamefont
  {A.}~\bibnamefont {Szameit}},\ }\href {\doibase 10.1038/nature12066}
  {\bibfield  {journal} {\bibinfo  {journal} {Nature}\ }\textbf {\bibinfo
  {volume} {496}},\ \bibinfo {pages} {196} (\bibinfo {year}
  {2013}{\natexlab{a}})}\BibitemShut {NoStop}%
\bibitem [{\citenamefont {Peleg}\ \emph {et~al.}(2007)\citenamefont {Peleg},
  \citenamefont {Bartal}, \citenamefont {Freedman}, \citenamefont {Manela},
  \citenamefont {Segev},\ and\ \citenamefont
  {Christodoulides}}]{peleg_prl_2007}%
  \BibitemOpen
  \bibfield  {author} {\bibinfo {author} {\bibfnamefont {O.}~\bibnamefont
  {Peleg}}, \bibinfo {author} {\bibfnamefont {G.}~\bibnamefont {Bartal}},
  \bibinfo {author} {\bibfnamefont {B.}~\bibnamefont {Freedman}}, \bibinfo
  {author} {\bibfnamefont {O.}~\bibnamefont {Manela}}, \bibinfo {author}
  {\bibfnamefont {M.}~\bibnamefont {Segev}}, \ and\ \bibinfo {author}
  {\bibfnamefont {D.~N.}\ \bibnamefont {Christodoulides}},\ }\href {\doibase
  10.1103/PhysRevLett.98.103901} {\bibfield  {journal} {\bibinfo  {journal}
  {Phys. Rev. Lett.}\ }\textbf {\bibinfo {volume} {98}},\ \bibinfo {pages}
  {103901} (\bibinfo {year} {2007})}\BibitemShut {NoStop}%
\bibitem [{\citenamefont {Ablowitz}\ \emph {et~al.}(2009)\citenamefont
  {Ablowitz}, \citenamefont {Nixon},\ and\ \citenamefont
  {Zhu}}]{ablowitz_pra_2009}%
  \BibitemOpen
  \bibfield  {author} {\bibinfo {author} {\bibfnamefont {M.~J.}\ \bibnamefont
  {Ablowitz}}, \bibinfo {author} {\bibfnamefont {S.~D.}\ \bibnamefont {Nixon}},
  \ and\ \bibinfo {author} {\bibfnamefont {Y.}~\bibnamefont {Zhu}},\ }\href
  {\doibase 10.1103/PhysRevA.79.053830} {\bibfield  {journal} {\bibinfo
  {journal} {Phys. Rev. A}\ }\textbf {\bibinfo {volume} {79}},\ \bibinfo
  {pages} {053830} (\bibinfo {year} {2009})}\BibitemShut {NoStop}%
\bibitem [{\citenamefont {Lee}\ \emph {et~al.}(2009)\citenamefont {Lee},
  \citenamefont {Gr\'emaud}, \citenamefont {Han}, \citenamefont {Englert},\
  and\ \citenamefont {Miniatura}}]{lee_pra_2009}%
  \BibitemOpen
  \bibfield  {author} {\bibinfo {author} {\bibfnamefont {K.~L.}\ \bibnamefont
  {Lee}}, \bibinfo {author} {\bibfnamefont {B.}~\bibnamefont {Gr\'emaud}},
  \bibinfo {author} {\bibfnamefont {R.}~\bibnamefont {Han}}, \bibinfo {author}
  {\bibfnamefont {B.-G.}\ \bibnamefont {Englert}}, \ and\ \bibinfo {author}
  {\bibfnamefont {C.}~\bibnamefont {Miniatura}},\ }\href {\doibase
  10.1103/PhysRevA.80.043411} {\bibfield  {journal} {\bibinfo  {journal} {Phys.
  Rev. A}\ }\textbf {\bibinfo {volume} {80}},\ \bibinfo {pages} {043411}
  (\bibinfo {year} {2009})}\BibitemShut {NoStop}%
\bibitem [{\citenamefont {Bahat-Treidel}\ \emph {et~al.}(2010)\citenamefont
  {Bahat-Treidel}, \citenamefont {Peleg}, \citenamefont {Grobman},
  \citenamefont {Shapira}, \citenamefont {Segev},\ and\ \citenamefont
  {Pereg-Barnea}}]{bahat-treidel_prl_2010}%
  \BibitemOpen
  \bibfield  {author} {\bibinfo {author} {\bibfnamefont {O.}~\bibnamefont
  {Bahat-Treidel}}, \bibinfo {author} {\bibfnamefont {O.}~\bibnamefont
  {Peleg}}, \bibinfo {author} {\bibfnamefont {M.}~\bibnamefont {Grobman}},
  \bibinfo {author} {\bibfnamefont {N.}~\bibnamefont {Shapira}}, \bibinfo
  {author} {\bibfnamefont {M.}~\bibnamefont {Segev}}, \ and\ \bibinfo {author}
  {\bibfnamefont {T.}~\bibnamefont {Pereg-Barnea}},\ }\href {\doibase
  10.1103/PhysRevLett.104.063901} {\bibfield  {journal} {\bibinfo  {journal}
  {Phys. Rev. Lett.}\ }\textbf {\bibinfo {volume} {104}},\ \bibinfo {pages}
  {063901} (\bibinfo {year} {2010})}\BibitemShut {NoStop}%
\bibitem [{\citenamefont {Bahat-Treidel}\ and\ \citenamefont
  {Segev}(2011)}]{bahat-treidel_pra_2011}%
  \BibitemOpen
  \bibfield  {author} {\bibinfo {author} {\bibfnamefont {O.}~\bibnamefont
  {Bahat-Treidel}}\ and\ \bibinfo {author} {\bibfnamefont {M.}~\bibnamefont
  {Segev}},\ }\href {\doibase 10.1103/PhysRevA.84.021802} {\bibfield  {journal}
  {\bibinfo  {journal} {Phys. Rev. A}\ }\textbf {\bibinfo {volume} {84}},\
  \bibinfo {pages} {021802} (\bibinfo {year} {2011})}\BibitemShut {NoStop}%
\bibitem [{\citenamefont {Terhalle}\ \emph {et~al.}(2008)\citenamefont
  {Terhalle}, \citenamefont {Richter}, \citenamefont {Desyatnikov},
  \citenamefont {Neshev}, \citenamefont {Krolikowski}, \citenamefont {Kaiser},
  \citenamefont {Denz},\ and\ \citenamefont {Kivshar}}]{terhalle_prl_2008}%
  \BibitemOpen
  \bibfield  {author} {\bibinfo {author} {\bibfnamefont {B.}~\bibnamefont
  {Terhalle}}, \bibinfo {author} {\bibfnamefont {T.}~\bibnamefont {Richter}},
  \bibinfo {author} {\bibfnamefont {A.~S.}\ \bibnamefont {Desyatnikov}},
  \bibinfo {author} {\bibfnamefont {D.~N.}\ \bibnamefont {Neshev}}, \bibinfo
  {author} {\bibfnamefont {W.}~\bibnamefont {Krolikowski}}, \bibinfo {author}
  {\bibfnamefont {F.}~\bibnamefont {Kaiser}}, \bibinfo {author} {\bibfnamefont
  {C.}~\bibnamefont {Denz}}, \ and\ \bibinfo {author} {\bibfnamefont {Y.~S.}\
  \bibnamefont {Kivshar}},\ }\href {\doibase 10.1103/PhysRevLett.101.013903}
  {\bibfield  {journal} {\bibinfo  {journal} {Phys. Rev. Lett.}\ }\textbf
  {\bibinfo {volume} {101}},\ \bibinfo {pages} {013903} (\bibinfo {year}
  {2008})}\BibitemShut {NoStop}%
\bibitem [{\citenamefont {Plotnik}\ \emph {et~al.}(2014)\citenamefont
  {Plotnik}, \citenamefont {Rechtsman}, \citenamefont {Song}, \citenamefont
  {Heinrich}, \citenamefont {Zeuner}, \citenamefont {Nolte}, \citenamefont
  {Lumer}, \citenamefont {Malkova}, \citenamefont {Xu}, \citenamefont
  {Szameit}, \citenamefont {Chen},\ and\ \citenamefont
  {Segev}}]{plotnik_nm_2014}%
  \BibitemOpen
  \bibfield  {author} {\bibinfo {author} {\bibfnamefont {Y.}~\bibnamefont
  {Plotnik}}, \bibinfo {author} {\bibfnamefont {M.~C.}\ \bibnamefont
  {Rechtsman}}, \bibinfo {author} {\bibfnamefont {D.}~\bibnamefont {Song}},
  \bibinfo {author} {\bibfnamefont {M.}~\bibnamefont {Heinrich}}, \bibinfo
  {author} {\bibfnamefont {J.~M.}\ \bibnamefont {Zeuner}}, \bibinfo {author}
  {\bibfnamefont {S.}~\bibnamefont {Nolte}}, \bibinfo {author} {\bibfnamefont
  {Y.}~\bibnamefont {Lumer}}, \bibinfo {author} {\bibfnamefont
  {N.}~\bibnamefont {Malkova}}, \bibinfo {author} {\bibfnamefont
  {J.}~\bibnamefont {Xu}}, \bibinfo {author} {\bibfnamefont {A.}~\bibnamefont
  {Szameit}}, \bibinfo {author} {\bibfnamefont {Z.~C.}\ \bibnamefont {Chen}}, \
  and\ \bibinfo {author} {\bibfnamefont {M.}~\bibnamefont {Segev}},\ }\href
  {\doibase 10.1038/nmat3783} {\bibfield  {journal} {\bibinfo  {journal} {Nat.
  Mater.}\ }\textbf {\bibinfo {volume} {13}},\ \bibinfo {pages} {57} (\bibinfo
  {year} {2014})}\BibitemShut {NoStop}%
\bibitem [{\citenamefont {Zhang}\ \emph {et~al.}(2011)\citenamefont {Zhang},
  \citenamefont {Wang}, \citenamefont {Nie}, \citenamefont {Li}, \citenamefont
  {Chen}, \citenamefont {Lu},\ and\ \citenamefont {Xiao}}]{zhang_prl_2011}%
  \BibitemOpen
  \bibfield  {author} {\bibinfo {author} {\bibfnamefont {Y.}~\bibnamefont
  {Zhang}}, \bibinfo {author} {\bibfnamefont {Z.}~\bibnamefont {Wang}},
  \bibinfo {author} {\bibfnamefont {Z.}~\bibnamefont {Nie}}, \bibinfo {author}
  {\bibfnamefont {C.}~\bibnamefont {Li}}, \bibinfo {author} {\bibfnamefont
  {H.}~\bibnamefont {Chen}}, \bibinfo {author} {\bibfnamefont {K.}~\bibnamefont
  {Lu}}, \ and\ \bibinfo {author} {\bibfnamefont {M.}~\bibnamefont {Xiao}},\
  }\href {\doibase 10.1103/PhysRevLett.106.093904} {\bibfield  {journal}
  {\bibinfo  {journal} {Phys. Rev. Lett.}\ }\textbf {\bibinfo {volume} {106}},\
  \bibinfo {pages} {093904} (\bibinfo {year} {2011})}\BibitemShut {NoStop}%
\bibitem [{\citenamefont {Zhang}\ \emph {et~al.}(2010)\citenamefont {Zhang},
  \citenamefont {Nie}, \citenamefont {Zhao}, \citenamefont {Li}, \citenamefont
  {Wang}, \citenamefont {Si},\ and\ \citenamefont
  {Xiao}}]{zhangyanpeng_oe_2010}%
  \BibitemOpen
  \bibfield  {author} {\bibinfo {author} {\bibfnamefont {Y.}~\bibnamefont
  {Zhang}}, \bibinfo {author} {\bibfnamefont {Z.}~\bibnamefont {Nie}}, \bibinfo
  {author} {\bibfnamefont {Y.}~\bibnamefont {Zhao}}, \bibinfo {author}
  {\bibfnamefont {C.}~\bibnamefont {Li}}, \bibinfo {author} {\bibfnamefont
  {R.}~\bibnamefont {Wang}}, \bibinfo {author} {\bibfnamefont {J.}~\bibnamefont
  {Si}}, \ and\ \bibinfo {author} {\bibfnamefont {M.}~\bibnamefont {Xiao}},\
  }\href {\doibase 10.1364/OE.18.010963} {\bibfield  {journal} {\bibinfo
  {journal} {Opt. Express}\ }\textbf {\bibinfo {volume} {18}},\ \bibinfo
  {pages} {10963} (\bibinfo {year} {2010})}\BibitemShut {NoStop}%
\bibitem [{\citenamefont {Zhang}\ \emph {et~al.}(2013)\citenamefont {Zhang},
  \citenamefont {Yuan}, \citenamefont {Zhang}, \citenamefont {Zheng},
  \citenamefont {Chen}, \citenamefont {Li}, \citenamefont {Wang},\ and\
  \citenamefont {Xiao}}]{zhang_lpl_2013}%
  \BibitemOpen
  \bibfield  {author} {\bibinfo {author} {\bibfnamefont {Y.}~\bibnamefont
  {Zhang}}, \bibinfo {author} {\bibfnamefont {C.}~\bibnamefont {Yuan}},
  \bibinfo {author} {\bibfnamefont {Y.}~\bibnamefont {Zhang}}, \bibinfo
  {author} {\bibfnamefont {H.}~\bibnamefont {Zheng}}, \bibinfo {author}
  {\bibfnamefont {H.}~\bibnamefont {Chen}}, \bibinfo {author} {\bibfnamefont
  {C.}~\bibnamefont {Li}}, \bibinfo {author} {\bibfnamefont {Z.}~\bibnamefont
  {Wang}}, \ and\ \bibinfo {author} {\bibfnamefont {M.}~\bibnamefont {Xiao}},\
  }\href {\doibase 10.1088/1612-2011/10/5/055406} {\bibfield  {journal}
  {\bibinfo  {journal} {Laser Phys. Lett.}\ }\textbf {\bibinfo {volume} {10}},\
  \bibinfo {pages} {055406} (\bibinfo {year} {2013})}\BibitemShut {NoStop}%
\bibitem [{\citenamefont {Artoni}\ and\ \citenamefont
  {La~Rocca}(2006)}]{artoni_prl_2006}%
  \BibitemOpen
  \bibfield  {author} {\bibinfo {author} {\bibfnamefont {M.}~\bibnamefont
  {Artoni}}\ and\ \bibinfo {author} {\bibfnamefont {G.~C.}\ \bibnamefont
  {La~Rocca}},\ }\href {\doibase 10.1103/PhysRevLett.96.073905} {\bibfield
  {journal} {\bibinfo  {journal} {Phys. Rev. Lett.}\ }\textbf {\bibinfo
  {volume} {96}},\ \bibinfo {pages} {073905} (\bibinfo {year}
  {2006})}\BibitemShut {NoStop}%
\bibitem [{\citenamefont {Zhang}\ \emph {et~al.}(2012)\citenamefont {Zhang},
  \citenamefont {Yao}, \citenamefont {Yuan}, \citenamefont {Li}, \citenamefont
  {Yuan}, \citenamefont {Feng}, \citenamefont {Jia},\ and\ \citenamefont
  {Zhang}}]{zhang_ieee_2012}%
  \BibitemOpen
  \bibfield  {author} {\bibinfo {author} {\bibfnamefont {Y.}~\bibnamefont
  {Zhang}}, \bibinfo {author} {\bibfnamefont {X.}~\bibnamefont {Yao}}, \bibinfo
  {author} {\bibfnamefont {C.}~\bibnamefont {Yuan}}, \bibinfo {author}
  {\bibfnamefont {P.}~\bibnamefont {Li}}, \bibinfo {author} {\bibfnamefont
  {J.}~\bibnamefont {Yuan}}, \bibinfo {author} {\bibfnamefont {W.}~\bibnamefont
  {Feng}}, \bibinfo {author} {\bibfnamefont {S.}~\bibnamefont {Jia}}, \ and\
  \bibinfo {author} {\bibfnamefont {Y.}~\bibnamefont {Zhang}},\ }\href
  {\doibase 10.1109/JPHOT.2012.2225609} {\bibfield  {journal} {\bibinfo
  {journal} {IEEE Photon. J.}\ }\textbf {\bibinfo {volume} {4}},\ \bibinfo
  {pages} {2057} (\bibinfo {year} {2012})}\BibitemShut {NoStop}%
\bibitem [{\citenamefont {Wang}\ \emph {et~al.}(2013)\citenamefont {Wang},
  \citenamefont {Zhou}, \citenamefont {Guo}, \citenamefont {Zhang},
  \citenamefont {Evers},\ and\ \citenamefont {Zhu}}]{wang_prl_2013}%
  \BibitemOpen
  \bibfield  {author} {\bibinfo {author} {\bibfnamefont {D.-W.}\ \bibnamefont
  {Wang}}, \bibinfo {author} {\bibfnamefont {H.-T.}\ \bibnamefont {Zhou}},
  \bibinfo {author} {\bibfnamefont {M.-J.}\ \bibnamefont {Guo}}, \bibinfo
  {author} {\bibfnamefont {J.-X.}\ \bibnamefont {Zhang}}, \bibinfo {author}
  {\bibfnamefont {J.}~\bibnamefont {Evers}}, \ and\ \bibinfo {author}
  {\bibfnamefont {S.-Y.}\ \bibnamefont {Zhu}},\ }\href {\doibase
  10.1103/PhysRevLett.110.093901} {\bibfield  {journal} {\bibinfo  {journal}
  {Phys. Rev. Lett.}\ }\textbf {\bibinfo {volume} {110}},\ \bibinfo {pages}
  {093901} (\bibinfo {year} {2013})}\BibitemShut {NoStop}%
\bibitem [{\citenamefont {Jotzu}\ \emph {et~al.}(2014)\citenamefont {Jotzu},
  \citenamefont {Messer}, \citenamefont {Desbuquois}, \citenamefont {Lebrat},
  \citenamefont {Uehlinger}, \citenamefont {Greif},\ and\ \citenamefont
  {Esslinger}}]{jotzu_nature_2014}%
  \BibitemOpen
  \bibfield  {author} {\bibinfo {author} {\bibfnamefont {G.}~\bibnamefont
  {Jotzu}}, \bibinfo {author} {\bibfnamefont {M.}~\bibnamefont {Messer}},
  \bibinfo {author} {\bibfnamefont {R.}~\bibnamefont {Desbuquois}}, \bibinfo
  {author} {\bibfnamefont {M.}~\bibnamefont {Lebrat}}, \bibinfo {author}
  {\bibfnamefont {T.}~\bibnamefont {Uehlinger}}, \bibinfo {author}
  {\bibfnamefont {D.}~\bibnamefont {Greif}}, \ and\ \bibinfo {author}
  {\bibfnamefont {T.}~\bibnamefont {Esslinger}},\ }\href {\doibase
  10.1038/nature13915} {\bibfield  {journal} {\bibinfo  {journal} {Nature}\
  }\textbf {\bibinfo {volume} {515}},\ \bibinfo {pages} {237} (\bibinfo {year}
  {2014})}\BibitemShut {NoStop}%
\bibitem [{\citenamefont {Pang}\ \emph {et~al.}(2011)\citenamefont {Pang},
  \citenamefont {Wu}, \citenamefont {Yuan}, \citenamefont {Liu},\ and\
  \citenamefont {Chen}}]{pang_jpsj_2011}%
  \BibitemOpen
  \bibfield  {author} {\bibinfo {author} {\bibfnamefont {W.}~\bibnamefont
  {Pang}}, \bibinfo {author} {\bibfnamefont {J.}~\bibnamefont {Wu}}, \bibinfo
  {author} {\bibfnamefont {Z.}~\bibnamefont {Yuan}}, \bibinfo {author}
  {\bibfnamefont {Y.}~\bibnamefont {Liu}}, \ and\ \bibinfo {author}
  {\bibfnamefont {G.}~\bibnamefont {Chen}},\ }\href {\doibase
  10.1143/JPSJ.80.113401} {\bibfield  {journal} {\bibinfo  {journal} {J. Phys.
  Soc. Jap.}\ }\textbf {\bibinfo {volume} {80}},\ \bibinfo {pages} {113401}
  (\bibinfo {year} {2011})}\BibitemShut {NoStop}%
\bibitem [{\citenamefont {Li}\ \emph {et~al.}(2010)\citenamefont {Li},
  \citenamefont {Malomed}, \citenamefont {Feng},\ and\ \citenamefont
  {Zhou}}]{liyongyao_pra_2010}%
  \BibitemOpen
  \bibfield  {author} {\bibinfo {author} {\bibfnamefont {Y.}~\bibnamefont
  {Li}}, \bibinfo {author} {\bibfnamefont {B.~A.}\ \bibnamefont {Malomed}},
  \bibinfo {author} {\bibfnamefont {M.}~\bibnamefont {Feng}}, \ and\ \bibinfo
  {author} {\bibfnamefont {J.}~\bibnamefont {Zhou}},\ }\href {\doibase
  10.1103/PhysRevA.82.063813} {\bibfield  {journal} {\bibinfo  {journal} {Phys.
  Rev. A}\ }\textbf {\bibinfo {volume} {82}},\ \bibinfo {pages} {063813}
  (\bibinfo {year} {2010})}\BibitemShut {NoStop}%
\bibitem [{\citenamefont {Li}\ \emph {et~al.}(2012)\citenamefont {Li},
  \citenamefont {Pang}, \citenamefont {Fu},\ and\ \citenamefont
  {Malomed}}]{liyongyao_pra_2012}%
  \BibitemOpen
  \bibfield  {author} {\bibinfo {author} {\bibfnamefont {Y.}~\bibnamefont
  {Li}}, \bibinfo {author} {\bibfnamefont {W.}~\bibnamefont {Pang}}, \bibinfo
  {author} {\bibfnamefont {S.}~\bibnamefont {Fu}}, \ and\ \bibinfo {author}
  {\bibfnamefont {B.~A.}\ \bibnamefont {Malomed}},\ }\href {\doibase
  10.1103/PhysRevA.85.053821} {\bibfield  {journal} {\bibinfo  {journal} {Phys.
  Rev. A}\ }\textbf {\bibinfo {volume} {85}},\ \bibinfo {pages} {053821}
  (\bibinfo {year} {2012})}\BibitemShut {NoStop}%
\bibitem [{\citenamefont {Wu}\ \emph {et~al.}(2008)\citenamefont {Wu},
  \citenamefont {Artoni},\ and\ \citenamefont {Rocca}}]{wu_josab_2008}%
  \BibitemOpen
  \bibfield  {author} {\bibinfo {author} {\bibfnamefont {J.-H.}\ \bibnamefont
  {Wu}}, \bibinfo {author} {\bibfnamefont {M.}~\bibnamefont {Artoni}}, \ and\
  \bibinfo {author} {\bibfnamefont {G.~C.~L.}\ \bibnamefont {Rocca}},\ }\href
  {\doibase 10.1364/JOSAB.25.001840} {\bibfield  {journal} {\bibinfo  {journal}
  {J. Opt. Soc. Am. B}\ }\textbf {\bibinfo {volume} {25}},\ \bibinfo {pages}
  {1840} (\bibinfo {year} {2008})}\BibitemShut {NoStop}%
\bibitem [{\citenamefont {Steck}()}]{steck_alkali}%
  \BibitemOpen
  \bibfield  {author} {\bibinfo {author} {\bibfnamefont {D.~A.}\ \bibnamefont
  {Steck}},\ }\href@noop {} {\bibinfo  {journal}
  {\url{http://steck.us/alkalidata}}\ }\BibitemShut {NoStop}%
\bibitem [{\citenamefont {Wu}\ \emph {et~al.}(2013)\citenamefont {Wu},
  \citenamefont {Zhang}, \citenamefont {Yuan}, \citenamefont {Wen},
  \citenamefont {Zheng}, \citenamefont {Zhang},\ and\ \citenamefont
  {Xiao}}]{wu_pra_2013}%
  \BibitemOpen
\bibfield  {journal} {  }\bibfield  {author} {\bibinfo {author} {\bibfnamefont
  {Z.}~\bibnamefont {Wu}}, \bibinfo {author} {\bibfnamefont {Y.}~\bibnamefont
  {Zhang}}, \bibinfo {author} {\bibfnamefont {C.}~\bibnamefont {Yuan}},
  \bibinfo {author} {\bibfnamefont {F.}~\bibnamefont {Wen}}, \bibinfo {author}
  {\bibfnamefont {H.}~\bibnamefont {Zheng}}, \bibinfo {author} {\bibfnamefont
  {Y.}~\bibnamefont {Zhang}}, \ and\ \bibinfo {author} {\bibfnamefont
  {M.}~\bibnamefont {Xiao}},\ }\href {\doibase 10.1103/PhysRevA.88.063828}
  {\bibfield  {journal} {\bibinfo  {journal} {Phys. Rev. A}\ }\textbf {\bibinfo
  {volume} {88}},\ \bibinfo {pages} {063828} (\bibinfo {year}
  {2013})}\BibitemShut {NoStop}%
\bibitem [{\citenamefont {Paredes}\ \emph {et~al.}(2014)\citenamefont
  {Paredes}, \citenamefont {Feijoo},\ and\ \citenamefont
  {Michinel}}]{paredes_prl_2014}%
  \BibitemOpen
  \bibfield  {author} {\bibinfo {author} {\bibfnamefont {A.}~\bibnamefont
  {Paredes}}, \bibinfo {author} {\bibfnamefont {D.}~\bibnamefont {Feijoo}}, \
  and\ \bibinfo {author} {\bibfnamefont {H.}~\bibnamefont {Michinel}},\ }\href
  {\doibase 10.1103/PhysRevLett.112.173901} {\bibfield  {journal} {\bibinfo
  {journal} {Phys. Rev. Lett.}\ }\textbf {\bibinfo {volume} {112}},\ \bibinfo
  {pages} {173901} (\bibinfo {year} {2014})}\BibitemShut {NoStop}%
\bibitem [{\citenamefont {Efremidis}\ \emph {et~al.}(2003)\citenamefont
  {Efremidis}, \citenamefont {Hudock}, \citenamefont {Christodoulides},
  \citenamefont {Fleischer}, \citenamefont {Cohen},\ and\ \citenamefont
  {Segev}}]{efremidis_prl_2003}%
  \BibitemOpen
  \bibfield  {author} {\bibinfo {author} {\bibfnamefont {N.~K.}\ \bibnamefont
  {Efremidis}}, \bibinfo {author} {\bibfnamefont {J.}~\bibnamefont {Hudock}},
  \bibinfo {author} {\bibfnamefont {D.~N.}\ \bibnamefont {Christodoulides}},
  \bibinfo {author} {\bibfnamefont {J.~W.}\ \bibnamefont {Fleischer}}, \bibinfo
  {author} {\bibfnamefont {O.}~\bibnamefont {Cohen}}, \ and\ \bibinfo {author}
  {\bibfnamefont {M.}~\bibnamefont {Segev}},\ }\href {\doibase
  10.1103/PhysRevLett.91.213906} {\bibfield  {journal} {\bibinfo  {journal}
  {Phys. Rev. Lett.}\ }\textbf {\bibinfo {volume} {91}},\ \bibinfo {pages}
  {213906} (\bibinfo {year} {2003})}\BibitemShut {NoStop}%
\bibitem [{\citenamefont {Szameit}\ \emph {et~al.}(2011)\citenamefont
  {Szameit}, \citenamefont {Rechtsman}, \citenamefont {Bahat-Treidel},\ and\
  \citenamefont {Segev}}]{szameit_pra_2011}%
  \BibitemOpen
  \bibfield  {author} {\bibinfo {author} {\bibfnamefont {A.}~\bibnamefont
  {Szameit}}, \bibinfo {author} {\bibfnamefont {M.~C.}\ \bibnamefont
  {Rechtsman}}, \bibinfo {author} {\bibfnamefont {O.}~\bibnamefont
  {Bahat-Treidel}}, \ and\ \bibinfo {author} {\bibfnamefont {M.}~\bibnamefont
  {Segev}},\ }\href {\doibase 10.1103/PhysRevA.84.021806} {\bibfield  {journal}
  {\bibinfo  {journal} {Phys. Rev. A}\ }\textbf {\bibinfo {volume} {84}},\
  \bibinfo {pages} {021806} (\bibinfo {year} {2011})}\BibitemShut {NoStop}%
\bibitem [{\citenamefont {Rechtsman}\ \emph
  {et~al.}(2013{\natexlab{b}})\citenamefont {Rechtsman}, \citenamefont
  {Plotnik}, \citenamefont {Zeuner}, \citenamefont {Song}, \citenamefont
  {Chen}, \citenamefont {Szameit},\ and\ \citenamefont
  {Segev}}]{rechtsman_prl_2013}%
  \BibitemOpen
  \bibfield  {author} {\bibinfo {author} {\bibfnamefont {M.~C.}\ \bibnamefont
  {Rechtsman}}, \bibinfo {author} {\bibfnamefont {Y.}~\bibnamefont {Plotnik}},
  \bibinfo {author} {\bibfnamefont {J.~M.}\ \bibnamefont {Zeuner}}, \bibinfo
  {author} {\bibfnamefont {D.}~\bibnamefont {Song}}, \bibinfo {author}
  {\bibfnamefont {Z.}~\bibnamefont {Chen}}, \bibinfo {author} {\bibfnamefont
  {A.}~\bibnamefont {Szameit}}, \ and\ \bibinfo {author} {\bibfnamefont
  {M.}~\bibnamefont {Segev}},\ }\href {\doibase 10.1103/PhysRevLett.111.103901}
  {\bibfield  {journal} {\bibinfo  {journal} {Phys. Rev. Lett.}\ }\textbf
  {\bibinfo {volume} {111}},\ \bibinfo {pages} {103901} (\bibinfo {year}
  {2013}{\natexlab{b}})}\BibitemShut {NoStop}%
\bibitem [{\citenamefont {Oka}\ and\ \citenamefont
  {Aoki}(2009)}]{takashi_prb_2009}%
  \BibitemOpen
  \bibfield  {author} {\bibinfo {author} {\bibfnamefont {T.}~\bibnamefont
  {Oka}}\ and\ \bibinfo {author} {\bibfnamefont {H.}~\bibnamefont {Aoki}},\
  }\href {\doibase 10.1103/PhysRevB.79.081406} {\bibfield  {journal} {\bibinfo
  {journal} {Phys. Rev. B}\ }\textbf {\bibinfo {volume} {79}},\ \bibinfo
  {pages} {081406} (\bibinfo {year} {2009})}\BibitemShut {NoStop}%
\bibitem [{\citenamefont {Fang}\ \emph {et~al.}(2013)\citenamefont {Fang},
  \citenamefont {Yu},\ and\ \citenamefont {Fan}}]{fang_oe_2013}%
  \BibitemOpen
  \bibfield  {author} {\bibinfo {author} {\bibfnamefont {K.}~\bibnamefont
  {Fang}}, \bibinfo {author} {\bibfnamefont {Z.}~\bibnamefont {Yu}}, \ and\
  \bibinfo {author} {\bibfnamefont {S.}~\bibnamefont {Fan}},\ }\href {\doibase
  10.1364/OE.21.018216} {\bibfield  {journal} {\bibinfo  {journal} {Opt.
  Express}\ }\textbf {\bibinfo {volume} {21}},\ \bibinfo {pages} {18216}
  (\bibinfo {year} {2013})}\BibitemShut {NoStop}%
\bibitem [{\citenamefont {Kennedy}\ \emph {et~al.}(2002)\citenamefont
  {Kennedy}, \citenamefont {Brett}, \citenamefont {Toader},\ and\ \citenamefont
  {John}}]{kennedy_nanolett_2002}%
  \BibitemOpen
  \bibfield  {author} {\bibinfo {author} {\bibfnamefont {S.~R.}\ \bibnamefont
  {Kennedy}}, \bibinfo {author} {\bibfnamefont {M.~J.}\ \bibnamefont {Brett}},
  \bibinfo {author} {\bibfnamefont {O.}~\bibnamefont {Toader}}, \ and\ \bibinfo
  {author} {\bibfnamefont {S.}~\bibnamefont {John}},\ }\href {\doibase
  10.1021/nl015635q} {\bibfield  {journal} {\bibinfo  {journal} {Nano Lett.}\
  }\textbf {\bibinfo {volume} {2}},\ \bibinfo {pages} {59} (\bibinfo {year}
  {2002})}\BibitemShut {NoStop}%
\bibitem [{\citenamefont {Pang}\ \emph {et~al.}(2005)\citenamefont {Pang},
  \citenamefont {Lee}, \citenamefont {Lee}, \citenamefont {Tam}, \citenamefont
  {Chan},\ and\ \citenamefont {Sheng}}]{pang_oe_2005}%
  \BibitemOpen
  \bibfield  {author} {\bibinfo {author} {\bibfnamefont {Y.~K.}\ \bibnamefont
  {Pang}}, \bibinfo {author} {\bibfnamefont {J.}~\bibnamefont {Lee}}, \bibinfo
  {author} {\bibfnamefont {H.}~\bibnamefont {Lee}}, \bibinfo {author}
  {\bibfnamefont {W.~Y.}\ \bibnamefont {Tam}}, \bibinfo {author} {\bibfnamefont
  {C.}~\bibnamefont {Chan}}, \ and\ \bibinfo {author} {\bibfnamefont
  {P.}~\bibnamefont {Sheng}},\ }\href {\doibase 10.1364/OPEX.13.007615}
  {\bibfield  {journal} {\bibinfo  {journal} {Opt. Express}\ }\textbf {\bibinfo
  {volume} {13}},\ \bibinfo {pages} {7615} (\bibinfo {year}
  {2005})}\BibitemShut {NoStop}%
\bibitem [{\citenamefont {Seet}\ \emph {et~al.}(2005)\citenamefont {Seet},
  \citenamefont {Mizeikis}, \citenamefont {Matsuo}, \citenamefont {Juodkazis},\
  and\ \citenamefont {Misawa}}]{seet_am_2005}%
  \BibitemOpen
  \bibfield  {author} {\bibinfo {author} {\bibfnamefont {K.~K.}\ \bibnamefont
  {Seet}}, \bibinfo {author} {\bibfnamefont {V.}~\bibnamefont {Mizeikis}},
  \bibinfo {author} {\bibfnamefont {S.}~\bibnamefont {Matsuo}}, \bibinfo
  {author} {\bibfnamefont {S.}~\bibnamefont {Juodkazis}}, \ and\ \bibinfo
  {author} {\bibfnamefont {H.}~\bibnamefont {Misawa}},\ }\href {\doibase
  10.1002/adma.200401527} {\bibfield  {journal} {\bibinfo  {journal} {Adv.
  Mater.}\ }\textbf {\bibinfo {volume} {17}},\ \bibinfo {pages} {541} (\bibinfo
  {year} {2005})}\BibitemShut {NoStop}%
\bibitem [{\citenamefont {Agrawal}(1990)}]{agrawal_prl_1990}%
  \BibitemOpen
  \bibfield  {author} {\bibinfo {author} {\bibfnamefont {G.~P.}\ \bibnamefont
  {Agrawal}},\ }\href {\doibase 10.1103/PhysRevLett.64.2487} {\bibfield
  {journal} {\bibinfo  {journal} {Phys. Rev. Lett.}\ }\textbf {\bibinfo
  {volume} {64}},\ \bibinfo {pages} {2487} (\bibinfo {year}
  {1990})}\BibitemShut {NoStop}%
\bibitem [{\citenamefont {Hickmann}\ \emph {et~al.}(1992)\citenamefont
  {Hickmann}, \citenamefont {Gomes},\ and\ \citenamefont
  {de~Ara\'ujo}}]{hickmann_prl_1992}%
  \BibitemOpen
  \bibfield  {author} {\bibinfo {author} {\bibfnamefont {J.~M.}\ \bibnamefont
  {Hickmann}}, \bibinfo {author} {\bibfnamefont {A.~S.~L.}\ \bibnamefont
  {Gomes}}, \ and\ \bibinfo {author} {\bibfnamefont {C.~B.}\ \bibnamefont
  {de~Ara\'ujo}},\ }\href {\doibase 10.1103/PhysRevLett.68.3547} {\bibfield
  {journal} {\bibinfo  {journal} {Phys. Rev. Lett.}\ }\textbf {\bibinfo
  {volume} {68}},\ \bibinfo {pages} {3547} (\bibinfo {year}
  {1992})}\BibitemShut {NoStop}%
\bibitem [{\citenamefont {Haldane}\ and\ \citenamefont
  {Raghu}(2008)}]{haldane_prl_2008}%
  \BibitemOpen
  \bibfield  {author} {\bibinfo {author} {\bibfnamefont {F.~D.~M.}\
  \bibnamefont {Haldane}}\ and\ \bibinfo {author} {\bibfnamefont
  {S.}~\bibnamefont {Raghu}},\ }\href {\doibase 10.1103/PhysRevLett.100.013904}
  {\bibfield  {journal} {\bibinfo  {journal} {Phys. Rev. Lett.}\ }\textbf
  {\bibinfo {volume} {100}},\ \bibinfo {pages} {013904} (\bibinfo {year}
  {2008})}\BibitemShut {NoStop}%
\bibitem [{\citenamefont {Wang}\ \emph {et~al.}(2009)\citenamefont {Wang},
  \citenamefont {Chong}, \citenamefont {Joannopoulos},\ and\ \citenamefont
  {Soljacic}}]{wang_nature_2009}%
  \BibitemOpen
  \bibfield  {author} {\bibinfo {author} {\bibfnamefont {Z.}~\bibnamefont
  {Wang}}, \bibinfo {author} {\bibfnamefont {Y.}~\bibnamefont {Chong}},
  \bibinfo {author} {\bibfnamefont {J.~D.}\ \bibnamefont {Joannopoulos}}, \
  and\ \bibinfo {author} {\bibfnamefont {M.}~\bibnamefont {Soljacic}},\ }\href
  {\doibase 10.1038/nature08293} {\bibfield  {journal} {\bibinfo  {journal}
  {Nature}\ }\textbf {\bibinfo {volume} {461}},\ \bibinfo {pages} {772}
  (\bibinfo {year} {2009})}\BibitemShut {NoStop}%
\bibitem [{\citenamefont {Zak}(1989)}]{zak_prl_1989}%
  \BibitemOpen
  \bibfield  {author} {\bibinfo {author} {\bibfnamefont {J.}~\bibnamefont
  {Zak}},\ }\href {\doibase 10.1103/PhysRevLett.62.2747} {\bibfield  {journal}
  {\bibinfo  {journal} {Phys. Rev. Lett.}\ }\textbf {\bibinfo {volume} {62}},\
  \bibinfo {pages} {2747} (\bibinfo {year} {1989})}\BibitemShut {NoStop}%
\bibitem [{\citenamefont {Mihalache}\ \emph {et~al.}(2003)\citenamefont
  {Mihalache}, \citenamefont {Mazilu}, \citenamefont {Towers}, \citenamefont
  {Malomed},\ and\ \citenamefont {Lederer}}]{mihalache_pre_2003}%
  \BibitemOpen
  \bibfield  {author} {\bibinfo {author} {\bibfnamefont {D.}~\bibnamefont
  {Mihalache}}, \bibinfo {author} {\bibfnamefont {D.}~\bibnamefont {Mazilu}},
  \bibinfo {author} {\bibfnamefont {I.}~\bibnamefont {Towers}}, \bibinfo
  {author} {\bibfnamefont {B.~A.}\ \bibnamefont {Malomed}}, \ and\ \bibinfo
  {author} {\bibfnamefont {F.}~\bibnamefont {Lederer}},\ }\href {\doibase
  10.1103/PhysRevE.67.056608} {\bibfield  {journal} {\bibinfo  {journal} {Phys.
  Rev. E}\ }\textbf {\bibinfo {volume} {67}},\ \bibinfo {pages} {056608}
  (\bibinfo {year} {2003})}\BibitemShut {NoStop}%
\bibitem [{\citenamefont {Desyatnikov}\ \emph {et~al.}(2005)\citenamefont
  {Desyatnikov}, \citenamefont {Mihalache}, \citenamefont {Mazilu},
  \citenamefont {Malomed}, \citenamefont {Denz},\ and\ \citenamefont
  {Lederer}}]{desyatnikov_pre_2005}%
  \BibitemOpen
  \bibfield  {author} {\bibinfo {author} {\bibfnamefont {A.~S.}\ \bibnamefont
  {Desyatnikov}}, \bibinfo {author} {\bibfnamefont {D.}~\bibnamefont
  {Mihalache}}, \bibinfo {author} {\bibfnamefont {D.}~\bibnamefont {Mazilu}},
  \bibinfo {author} {\bibfnamefont {B.~A.}\ \bibnamefont {Malomed}}, \bibinfo
  {author} {\bibfnamefont {C.}~\bibnamefont {Denz}}, \ and\ \bibinfo {author}
  {\bibfnamefont {F.}~\bibnamefont {Lederer}},\ }\href {\doibase
  10.1103/PhysRevE.71.026615} {\bibfield  {journal} {\bibinfo  {journal} {Phys.
  Rev. E}\ }\textbf {\bibinfo {volume} {71}},\ \bibinfo {pages} {026615}
  (\bibinfo {year} {2005})}\BibitemShut {NoStop}%
\bibitem [{\citenamefont {Michinel}\ \emph {et~al.}(2006)\citenamefont
  {Michinel}, \citenamefont {Paz-Alonso},\ and\ \citenamefont
  {P{\'e}rez-Garc{\'i}a}}]{michinel_prl_2006}%
  \BibitemOpen
  \bibfield  {author} {\bibinfo {author} {\bibfnamefont {H.}~\bibnamefont
  {Michinel}}, \bibinfo {author} {\bibfnamefont {M.~J.}\ \bibnamefont
  {Paz-Alonso}}, \ and\ \bibinfo {author} {\bibfnamefont {V.~M.}\ \bibnamefont
  {P{\'e}rez-Garc{\'i}a}},\ }\href {\doibase 10.1103/PhysRevLett.96.023903}
  {\bibfield  {journal} {\bibinfo  {journal} {Phys. Rev. Lett.}\ }\textbf
  {\bibinfo {volume} {96}},\ \bibinfo {pages} {023903} (\bibinfo {year}
  {2006})}\BibitemShut {NoStop}%
\bibitem [{\citenamefont {Hang}\ \emph {et~al.}(2013)\citenamefont {Hang},
  \citenamefont {Huang},\ and\ \citenamefont {Konotop}}]{hang_prl_2013}%
  \BibitemOpen
  \bibfield  {author} {\bibinfo {author} {\bibfnamefont {C.}~\bibnamefont
  {Hang}}, \bibinfo {author} {\bibfnamefont {G.}~\bibnamefont {Huang}}, \ and\
  \bibinfo {author} {\bibfnamefont {V.~V.}\ \bibnamefont {Konotop}},\ }\href
  {\doibase 10.1103/PhysRevLett.110.083604} {\bibfield  {journal} {\bibinfo
  {journal} {Phys. Rev. Lett.}\ }\textbf {\bibinfo {volume} {110}},\ \bibinfo
  {pages} {083604} (\bibinfo {year} {2013})}\BibitemShut {NoStop}%
\bibitem [{\citenamefont {Sheng}\ \emph {et~al.}(2013)\citenamefont {Sheng},
  \citenamefont {Miri}, \citenamefont {Christodoulides},\ and\ \citenamefont
  {Xiao}}]{sheng_pra_2013}%
  \BibitemOpen
  \bibfield  {author} {\bibinfo {author} {\bibfnamefont {J.}~\bibnamefont
  {Sheng}}, \bibinfo {author} {\bibfnamefont {M.-A.}\ \bibnamefont {Miri}},
  \bibinfo {author} {\bibfnamefont {D.~N.}\ \bibnamefont {Christodoulides}}, \
  and\ \bibinfo {author} {\bibfnamefont {M.}~\bibnamefont {Xiao}},\ }\href
  {\doibase 10.1103/PhysRevA.88.041803} {\bibfield  {journal} {\bibinfo
  {journal} {Phys. Rev. A}\ }\textbf {\bibinfo {volume} {88}},\ \bibinfo
  {pages} {041803} (\bibinfo {year} {2013})}\BibitemShut {NoStop}%
\bibitem [{\citenamefont {Sun}\ \emph {et~al.}(2011)\citenamefont {Sun},
  \citenamefont {Gu}, \citenamefont {Katsura},\ and\ \citenamefont
  {Das~Sarma}}]{sun_prl_2011}%
  \BibitemOpen
  \bibfield  {author} {\bibinfo {author} {\bibfnamefont {K.}~\bibnamefont
  {Sun}}, \bibinfo {author} {\bibfnamefont {Z.}~\bibnamefont {Gu}}, \bibinfo
  {author} {\bibfnamefont {H.}~\bibnamefont {Katsura}}, \ and\ \bibinfo
  {author} {\bibfnamefont {S.}~\bibnamefont {Das~Sarma}},\ }\href {\doibase
  10.1103/PhysRevLett.106.236803} {\bibfield  {journal} {\bibinfo  {journal}
  {Phys. Rev. Lett.}\ }\textbf {\bibinfo {volume} {106}},\ \bibinfo {pages}
  {236803} (\bibinfo {year} {2011})}\BibitemShut {NoStop}%
\bibitem [{\citenamefont {Crespi}\ \emph {et~al.}(2013)\citenamefont {Crespi},
  \citenamefont {Corrielli}, \citenamefont {Valle}, \citenamefont {Osellame},\
  and\ \citenamefont {Longhi}}]{crespi_njp_2013}%
  \BibitemOpen
  \bibfield  {author} {\bibinfo {author} {\bibfnamefont {A.}~\bibnamefont
  {Crespi}}, \bibinfo {author} {\bibfnamefont {G.}~\bibnamefont {Corrielli}},
  \bibinfo {author} {\bibfnamefont {G.~D.}\ \bibnamefont {Valle}}, \bibinfo
  {author} {\bibfnamefont {R.}~\bibnamefont {Osellame}}, \ and\ \bibinfo
  {author} {\bibfnamefont {S.}~\bibnamefont {Longhi}},\ }\href {\doibase
  10.1088/1367-2630/15/1/013012} {\bibfield  {journal} {\bibinfo  {journal}
  {New J. Phys.}\ }\textbf {\bibinfo {volume} {15}},\ \bibinfo {pages} {013012}
  (\bibinfo {year} {2013})}\BibitemShut {NoStop}%
\end{thebibliography}
%

\end{document}